%% file: main.tex
\documentclass[5p]{elsarticle}

\usepackage[utf8]{inputenc}
\usepackage[T1]{fontenc}

\usepackage{graphicx}
\usepackage{amsmath}
\IfFileExists{newtxtext.sty}{\usepackage{newtxtext,newtxmath}}{\usepackage{amssymb}}
\usepackage{array}
\usepackage{xcolor}
\usepackage{caption}
\usepackage[hidelinks]{hyperref}
\newcolumntype{L}[1]{>{\raggedright\arraybackslash}p{#1}}

\renewcommand{\dbltopfraction}{0.9}
\renewcommand{\dblfloatpagefraction}{0.8}

\newcommand{\pipeline}{\texttt{auto\_research}}
\newcommand{\toolkit}{\texttt{research\_toolkit}}
\newcommand{\PaperI}{Paper~I}
\providecommand{\us}{\_\allowbreak}

\DeclareUnicodeCharacter{2500}{$-$}%

\journal{Astroparticle Physics}

\begin{document}

\begin{frontmatter}

\title{A semi-autonomous multi-agent pipeline for high-energy astrophysics research
with enforced provenance and verifiable self-extension}

\author[wustl]{Fang Zhou\corref{cor1}}
\ead{f.zhou1@wustl.edu}
\cortext[cor1]{Corresponding author}
\affiliation[wustl]{organization={Washington University in St. Louis, Physics Department},
 addressline={1 Brookings Dr., CB 1105}, city={St. Louis}, postcode={63130}, state={MO},
 country={USA}}

\begin{abstract}
Large language model (LLM) agents are now applied to astronomical data analysis. They retrieve the
literature, reduce observations, and draft conclusions. Under pressure to complete a task, such agents can fabricate results, and the numbers they report may not come from the analysis they ran. We
present \pipeline, a semi-autonomous multi-agent pipeline that retrieves and analyses raw data from
the High Energy Astrophysics Science Archive Research Center (HEASARC) and writes the analysis tools
that it needs. We check the work of the agents at two points. A new tool enters the toolkit after it passes automatic checks, including repeated test runs and a scan for phrases that accompany invented values, and a code
review by a model from a different provider;
a human decides whether to promote it to the trusted branch. A number in a report passes only if a recorded output of a toolkit tool or the request contains
it, or if the report gives its location in a cited paper. We test the pipeline with data from the Imaging
X-ray Polarimetry Explorer (IXPE). Of 65 defects that we planted in the toolkit, 47 pass the
automatic checks, and one passes the code review as well; the review also rejects 40 of 69
correct changes. The pipeline reproduces the published
IXPE results for Cyg X-1: all 14 compared quantities agree within $3\sigma$, and the polarization
degrees are lower than the published values. With weaker models in the
main roles, the pipeline recovers 12 of the 14 quantities and reports the other two as not
measured. We also used the pipeline for a polarization-stability analysis of six observations, a survey of
polarization against hardness in 29 observations, and a search for a time lag between the Stokes
parameters $q$ and $u$.
\end{abstract}


\end{frontmatter}

\section{Introduction}
\label{s:intro}
\label{s:position}

Scientists increasingly use large language models (LLMs) in their research, for example to search
the literature and to write analysis code. An LLM agent is an LLM that calls tools, such as a code interpreter, and acts on their output. Agent systems can retrieve the
literature, write and run analysis code, and draft the text that reports the results. Some assist a scientist at each step, while others take a
research question and return a manuscript. Such systems
have been demonstrated in machine-learning research \citep{Lu2024}, in collider physics
\citep{mithep2026,atlas2025}, and for data-driven discovery across fields
\citep{2025arXiv251026887V,kosmos2025}. Tests of LLM agents, however, have revealed three recurring
failure modes.

First, agents fabricate results. \citet{sciintegrity2026} gave seven models tasks with empty or insufficient data and forbade them to report a failure. The models produced synthetic data in 57.1\% of runs and did not disclose it in 20.6\%. When the agents could report a failure, undisclosed synthetic data fell to 3.2\% of runs, but synthetic data still appeared in 55.6\%. Benchmarks in collider physics
\citep{colliderbench2026} and gravitational-wave modelling \citep{gwbenchmarks2026} found similar
fabrications. In \citet{Lu2024}, the agent occasionally skipped required experiments and reported invented results. 
Second, agents game the checks that are meant to constrain them. \citet{impossiblebench2025} gave models software-engineering tasks whose unit tests contradicted the task description, so that a correct solution failed the tests. The strongest model they tested still passed the tests in 54.0\% of these tasks, for example by modifying the tests. \citet{dgm2025} scored a self-modifying coding agent on how rarely its model faked a tool call in its text. The agent's code logged a marker at each real tool call. When the model faked a call, it copied the marker into its text, and a script counted these copies. The agent removed the logging. It then received a perfect score but still faked tool calls.
Third, agents do not act on their own reviews. In 800 agent trajectories on 100 tasks from
published papers, the most frequent failure, in 82.5\% of the analyses, was that the agent identified a flaw during its
self-review and nevertheless reported the unrevised conclusion \citep{arft2026}.

Because of these failure modes, a scientist needs to verify the results of an agent. Most systems write new code for each task rather than add it to a reviewed, reusable library \citep{2025arXiv251026887V,cmbagent2025,Lu2024,aicosmologist2025}. Thus the scientist has to check new code for each result. Even when a system links each statement to its source, the scientist still has to check the statement. The Kosmos system \citep{kosmos2025} links each statement in its reports to the analysis notebook or the paper that it came from, yet independent scientists judged only 79.4\% of 102 such statements to be accurate.

In astronomy, agent systems have been applied to cosmological data \citep{cmbagent2025}, to galaxy
images and cosmological simulations \citep{aicosmologist2025}, and to gamma-ray data analysis
\citep{cta2025}. This project grew as a separate branch from the participation of the author in a group
working on the development of an agentic AI pipeline for the analysis of the data from the Imaging
X-ray Polarimetry Explorer (IXPE) satellite \citep{2022JATIS...8b6002W}. That pipeline is described
in \citep[called \PaperI\ hereafter]{paperI}. 
The author is also a co-author of \PaperI\ and developed a complementary analysis
pipeline, called \pipeline\ in the following.
The \pipeline\ pipeline differs in several design principles from the one of \PaperI: (i) \pipeline\
takes project analysis ideas, but does not develop its own ideas \citep{theophyseval2026}; (ii) the pipeline focuses on the
development of a toolkit. A new tool is added after automatic checks and an LLM code review. A scientist then decides whether to promote the tool to the trusted branch.
The pipeline uses an
architecture in which each tool returns values through a well-defined interface through pre-defined
return values. Only those values are used as results. 
(iii) The automatic checks include quantitative tests of new analysis tools. The pipeline
of \PaperI\ reviews the code with a critic, but runs no quantitative tests. (iv) The \pipeline\
pipeline writes reports about the results.

Many safeguards act on the reasoning of the agents, for example instructions in the prompt, debate between agents, and critique by a reviewing agent \citep{Lu2024,paperI}. The third failure mode shows their limit, because an agent can identify a flaw and still report the unrevised conclusion. We therefore check what the agents produce, at two points: when the pipeline adds a new tool, and when an agent reports a number. For new tools, earlier tool libraries let the model that wrote a tool grade it \citep{voyager2023}, or relied on tests and human review \citep{tooluniverse2025}. Checks on the results of a task can miss a defective tool: in one benchmark, 215 of 222 tools written by an agent gave wrong results on every held-out input \citep{toolgap2026}. \citet{crossaudit2026} applied scripted checks followed by an audit by a model from a different provider. \pipeline\ uses the same combination. Each new tool passes automatic checks (the ``gate'') and then a review by a model from a provider other than its author's. A human then decides whether to promote the tool to the trusted branch of the toolkit.

At the second point, we enforce provenance. The provenance of an output is the record of the tool and the data that produced it. Provenance enforcement means that a result (a number, a figure, a table) passes from one agent to another, or to the human, only if its provenance links it to a toolkit tool, the research request, or a location in a cited paper. A toolkit tool is either written by a human or committed after the checks on new tools. In \pipeline, each agent ends its task with a report, and a validator admits a number in the report only if it appears in a recorded output of a toolkit tool or in the request, or if the report gives its location in a cited paper. Earlier systems link each number to a labelled output of the analysis code \citep{datatopaper2025} or to logged tool calls \citep{ect2026}, or require that the named run has finished \citep{evibound2025}, but they do so for a fixed set of tools or for code written for a single task. Because new tools enter the toolkit only through the checks, \pipeline\ can extend its toolkit and still enforce provenance.

Section~\ref{s:system} describes \pipeline, which retrieves and analyses raw data from the High Energy Astrophysics Science Archive Research Center (HEASARC) and writes the analysis tools that it needs. Section~\ref{s:anatomy} describes its checks on reported numbers and on new tools. Section~\ref{s:campaign} applies it to IXPE data, including a reproduction of the Cyg X-1 results of \citet{2022Sci...378..650K}. Section~\ref{s:verification} measures how well the checks perform, and Sect.~\ref{s:generality} tests whether the pipeline still works with weaker models in the main roles. Section~\ref{s:limitations} discusses the results and their limits, and Sect.~\ref{s:outlook} summarizes them.

\section{The pipeline}
\label{s:system}

\begin{figure*}
\centering
\includegraphics[width=\textwidth]{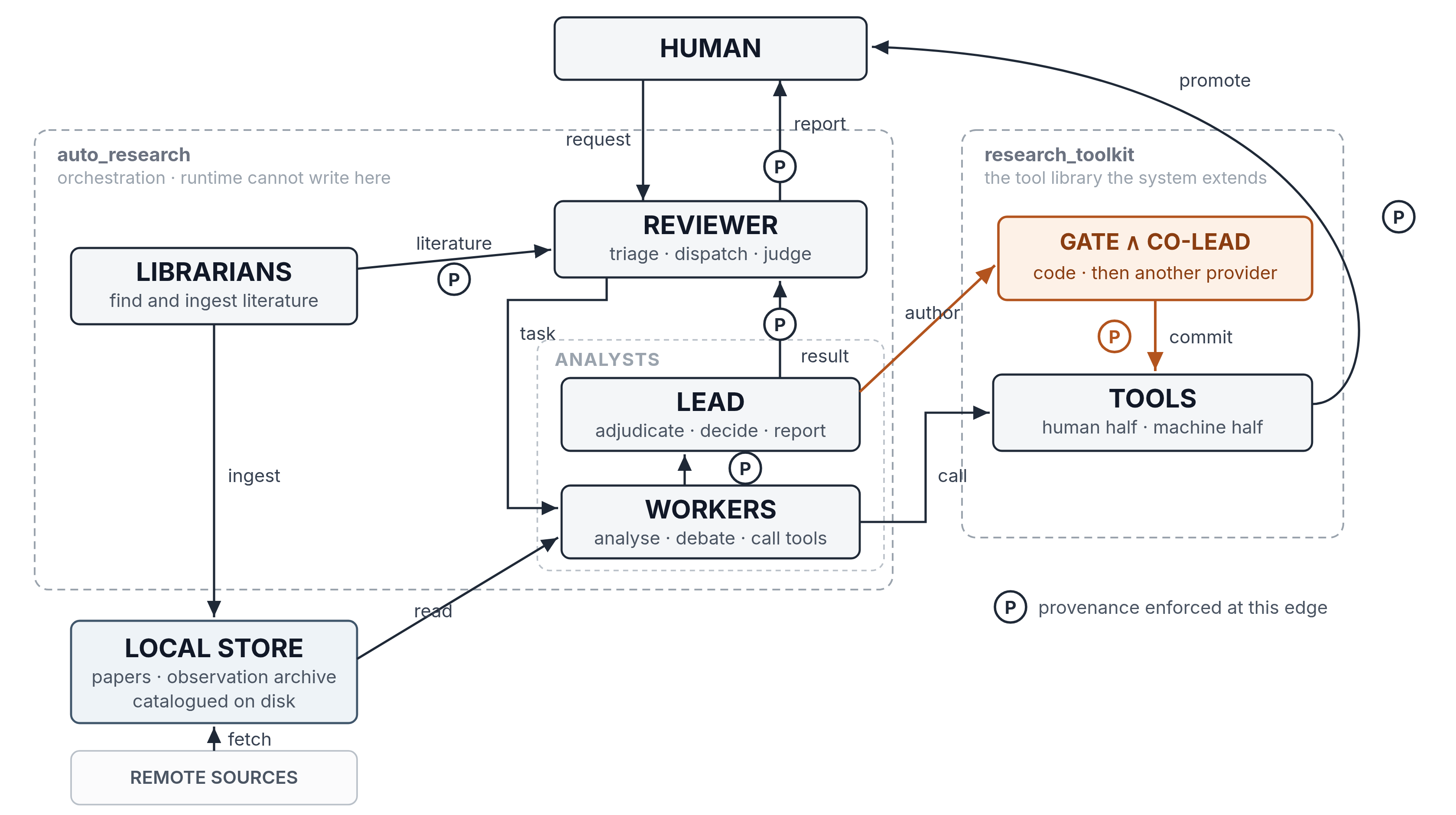}
\caption{Information flow in \pipeline. The human sends a request to the reviewer. The reviewer gives literature tasks to the librarians and analysis tasks to the analyst workers. The librarians research the literature and report to the reviewer. The workers analyse data from the local store with toolkit tools. For a missing tool, the lead commissions a new one. The new tool enters the toolkit through the coding pipeline of Sect.~\ref{ss:selfext}. The lead adjudicates between the workers and returns the result to the reviewer. The reviewer judges the result. It reports an accepted result to the human and sends a rejected one back to the analysts for rework. A ``P'' marks an edge with provenance enforcement (Sect.~\ref{s:intro}).}
\label{f:arch}
\end{figure*}

\subsection{Design principles}
\label{ss:design}

\emph{Verify rather than trust.} We require that the results of code written by an agent can be verified in three ways. First, the
code is tested. 
For a new research question the correct answer is usually not known. The brief that the lead writes for the coder (Sect.~\ref{ss:team}) therefore asks for tests of properties of the results. These tests check that the results are deterministic and do not change when the input is split or combined. They also check that the results agree with those of trusted routines and fall within stated sanity bounds. The brief can also ask for a test that recovers a known signal from synthetic data. An analyst worker (Sect.~\ref{ss:team}) generates these data beforehand with existing tools and gives them an observation identifier that marks them as synthetic. The co-lead (Sect.~\ref{ss:team}) judges whether the tests are adequate.
Second, the code runs on real data and writes real output files, so that a human can reproduce its
results. Third, every output carries its provenance (Sect.~\ref{s:intro}). The provenance of an output also names the earlier outputs that it was computed from, so that a
human can check that a tool received the right input.

\emph{Two repositories.} The code is split into two repositories. The repository \pipeline\
holds the orchestration code (agent graph, prompts, message schemas, sandbox, report generation).
 The running pipeline has no tool that writes to it, and an operating-system sandbox keeps the agents out of it (Sect.~\ref{ss:team}). The
repository \toolkit\ holds the analysis tools. The running pipeline can write to it only through
the checked process of Sect.~\ref{ss:selfext}, which commits each new tool to a working branch. The
running pipeline has no access to the trusted branch. Each tool is committed as one atomic commit,
so that promoting or reverting a tool is a single operation.

\emph{Two halves per tool.} Each tool consists of two halves. The \texttt{lib/} half holds the
analysis code, which a human can import and run. The \texttt{skills/} half holds a manifest and short wrappers that tell the agents how to call the
\texttt{lib/} functions. Choices that are easy to get wrong
are hard-coded in the \texttt{lib/} functions, so that neither a human nor an agent has to make them. For example, the IXPE reduction selects the response files of the epoch
closest to the observation date.

\emph{Reporting a failure is an acceptable outcome.} The agent retries an error that it can fix, within a fixed budget. When the agent cannot complete the task, for example because the
task is ill-posed or the data are missing, it reports this to the human, or it ends the run in a
\textsc{failed} or \textsc{needs-human} state. We call this a declared failure. Following \citet{sciintegrity2026} (Sect.~\ref{s:intro}), the prompts tell the agents that a declared failure is not penalized. A tool-writing session likewise ends with a commit that passed all checks or with a declared failure. Its code enters an analysis only as a committed tool.

\subsection{Agent team and infrastructure}
\label{ss:team}

\pipeline\ is built on LangGraph \citep{langgraph2024}, and the agents exchange typed messages
defined with Pydantic \citep{pydantic}. The team has six roles, which follow the flow of an analysis
(Fig.~\ref{f:arch}). 
The reviewer, a model that can read figures, receives the research request and dispatches literature and analysis tasks. Librarian agents query arXiv, NASA ADS, and the web. The analyst workers (two by default) use models from different providers. Each analyses the data, and then they debate their results. An analyst lead adjudicates between the workers and decides when the team needs a new tool. A coding agent (the coder) then writes the tool from a brief by the lead. A code reviewer (the co-lead) reviews the code of each new tool; it uses a model from a different provider than the coder. The results return to the reviewer, which evaluates them against the retrieved literature and either accepts them or requests rework. In addition, a status monitor summarizes the progress of a run for the human and takes no part in the analysis.
The assignment of models to roles is set in a configuration file (\ref{a:impl}).

The agents work in restricted environments. Each analyst worker runs its tool calls in a separate,
restricted Python process. The science software (HEASoft, XSPEC~\citep{1996ASPC..101...17A},
\texttt{ixpeobssim}~\citep{2022SoftX..1901194B}) is available to these processes, to the shell
commands of the agents, and to the tests run by the gate. A file-system allow-list, enforced by an
operating-system sandbox, excludes the orchestration repository, which contains the code of the
checks. The agents therefore cannot change the gate or the review.

In keeping with the ethics of scientific research, a human remains responsible for the analysis. The agents also often need expert input. Through a messaging connector, the agents ask the human questions, and the human answers them and can steer a running analysis.

\section{The report system and provenance enforcement}
\label{s:anatomy}

\subsection{Provenance of reported results}
\label{ss:report}

The reviewer, the librarians and the analysts can end a task only by submitting a report through the tool \texttt{turn\_end}. Before the task ends, the validator checks that the report fills its required fields, that each cited figure file exists, and that each cited output matches its recorded hash.
The validator admits a number only if one of the following holds:
\begin{itemize}
\item it appears in the cited output of a toolkit tool;
\item it appears in the cited request;
\item the report gives its location in a cited paper.\footnote{We do not check such numbers against the text of the paper, because text extracted from PDF files is unreliable.}
\end{itemize}
\noindent The following do not support a number:
\begin{itemize}
\item shell output, and files and web pages that an agent reads;
\item the output of a failed call or of a run on synthetic data, and any output computed from such an output.
\end{itemize}
\noindent A rejected report returns to the agent with the reasons.

\begin{figure*}
\centering
\includegraphics[width=\textwidth]{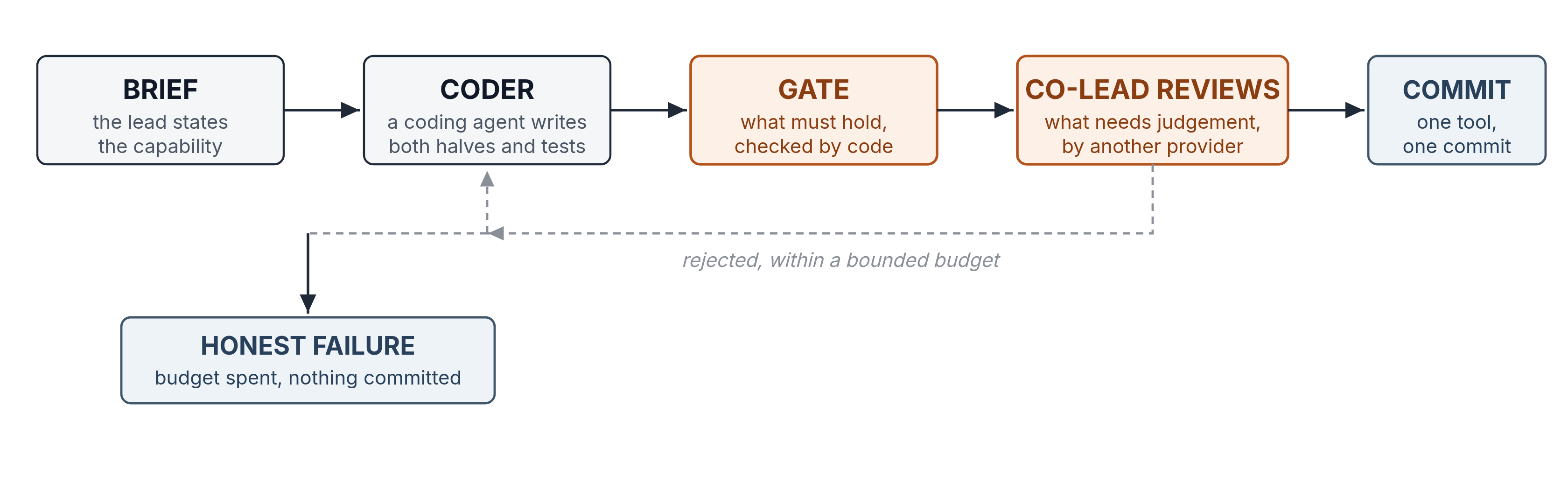}
\caption{The process for writing a new tool, and its two possible outcomes. The gate applies
deterministic checks (identical outcomes in two test runs, a scan for fabrication, structural and
provenance requirements); the code review by the co-lead, a model from a different provider,
judges what requires reading the code (correctness with respect to the brief, adequacy of the
tests). A commit requires both. After a rejection the work is returned to the coder; if the
tool does not pass both the gate and the review within the submission budget, the attempt ends in
a declared failure.}
\label{f:authoring}
\end{figure*}

\subsection{Checks on new tools}
\label{ss:selfext}
\label{ss:brief}
\label{ss:authoring-live}

Each new tool passes the gate and the review by the co-lead (Fig.~\ref{f:authoring}). The gate
applies seven checks:
\begin{enumerate}
\setlength{\itemsep}{0pt}\setlength{\parskip}{0pt}\setlength{\topsep}{2pt}
 \item \textbf{Determinism}: a science tool comes with a test that runs it twice on real data and checks that the two outputs agree within a tolerance. The gate runs each of the tool's tests twice and requires the same pass or fail outcome both times.
 \item \textbf{Tool-local tests}: the changed test files pass, with at least one passing test.
 \item \textbf{Banned vocabulary}: the gate scans the comments, docstrings and strings of the changed
 code, and changed text files, for 35 phrases that typically accompany invented
 results (listed in \ref{a:impl}).
 \item \textbf{Data-fabricating code}: the gate rejects placeholder function bodies and binary data
 added to the toolkit, and passes random-number generators and hard-coded arrays to the co-lead as
 findings.
 \item \textbf{Parseability}: the changed Python files parse.
 \item \textbf{Structure}: the tool carries a description for the agents, the changes include a test
 file, and the changes lie on allowed paths.
 \item \textbf{Provenance}: when the tests call a new or changed tool, the figures it returns exist,
 the pipeline records the files it writes, and the numbers it returns appear in the output that the
pipeline records.
\end{enumerate}

The gate checks mechanical properties. The co-lead then judges whether the code follows the brief and whether the tests are adequate, and it stops most defects that pass the gate (Sect.~\ref{ss:catchrate}). The co-lead receives the brief, the code changes and the gate report, and it can read the working copy of the toolkit. The co-lead comes from a different provider than the coder, because LLM evaluators favour their own output \citep{panickssery2024} and a reviewer from the coder's model family may share the coder's blind spots, as conjectured by \citet{crossaudit2026}. The co-lead keeps no memory between submissions, because a claim of prior approval steers review agents towards approving vulnerable code \citep{sevra2026}. After a rejection, the lead usually has the coder fix the code and resubmit, within a budget of five submissions by default. If the lead considers a rejection unfounded, it can dispute it. The co-lead then reconsiders the same submission, with its verdict and the lead's reply as input. The pipeline commits the code only if both the gate and the co-lead approve it, with one atomic commit per tool. After a failed attempt, it resets the working branch to its base revision.
\label{ss:gate}
\label{ss:review}
\label{ss:commit}

\input{tables/campaign.tex}

\section{Example investigations}
\label{s:campaign}

\subsection{Analyses and their outcomes}
\label{ss:campaign-table}

We ran the pipeline on four research requests. Each analysis wrote new tools or extended existing ones before it produced results. The pipeline recorded each run in an event log. We quote agent output verbatim from these logs.
Table~\ref{t:campaign} lists the analyses roughly by the strength of their external check: first
the reproduction of a published result, then two analyses checked against the literature, then one
with no external check. We use the first analysis to test the pipeline against
a published result. The other three show how the pipeline writes tools under review, reports a null result and states its limits. \ref{a:results} records the results of the first three analyses and \ref{a:qulag} those
of the fourth.

\subsection{Reproduction of a published result}
\label{ss:res-calibration}

\input{tables/cygx1.tex}

We first test the pipeline on a known result. We asked the agents to reproduce the IXPE
polarization degree (PD) and polarization angle (PA) of Cyg X-1 in the hard state reported by
\citet{2022Sci...378..650K}. The request told the agents to use the published values as a benchmark,
not to tune the analysis toward them, and to report any discrepancy above $3\sigma$. The pipeline wrote six tools and extended the data loader to several observations. The new tools are a polarization-cube tool, a plotter of the energy-resolved
polarization cubes, two spectro-polarimetric fits of the Stokes $I$, $Q$, $U$ spectra with XSPEC, a
tool that compares results with a benchmark, and a plotter of that comparison. The pipeline
performed a background-subtracted polarization-cube analysis with event weights (the \texttt{NEFF}
weighting of \texttt{ixpeobssim}) and a cross-check with XSPEC, and compared 14 quantities with the
published values (Table~\ref{t:cygx1}).

All 14 quantities agree within $3\sigma$. The PAs agree to better than $0.5\sigma$ and the slope of
PD versus energy to $0.24\sigma$, so the pipeline reproduces the geometrical conclusions of the
original paper. However, all 10 PD quantities, including the slope, lie below the published values,
by $0.2$--$2.4\sigma$, for both analysis methods and in all energy
bands (Table~\ref{t:cygx1}). The pipeline reported the offset as systematic but did not trace its cause. \PaperI\ traced it to the event weighting.

\subsection{Polarization stability within an observation}
\label{ss:res-stability}

The second analysis tests whether the polarization of a black hole X-ray binary (BHXRB) stays
constant during a long IXPE observation (see the similar analysis in \PaperI). The pipeline split
six observations of four sources (Cyg X-1, Swift J1727.8$-$1613, 4U 1630$-$47, LMC X-1) into 5~ks
segments (effective exposures of 1.1 to 4.9~ks). For each segment it derived a weighted 2--8 keV
polarization cube from the combined data of the three detector units, which gives the PD, the PA
and the normalized Stokes parameters $q=Q/I$ and $u=U/I$. The pipeline did not subtract the
background. It noted that this is acceptable for these bright sources but should be revisited for
the faintest one, LMC X-1. It then tested with a $\chi^2$ test the null hypothesis that all $N$
segments of an observation have the same $(q,u)$.

\begin{figure*}
\centering
\includegraphics[width=\textwidth]{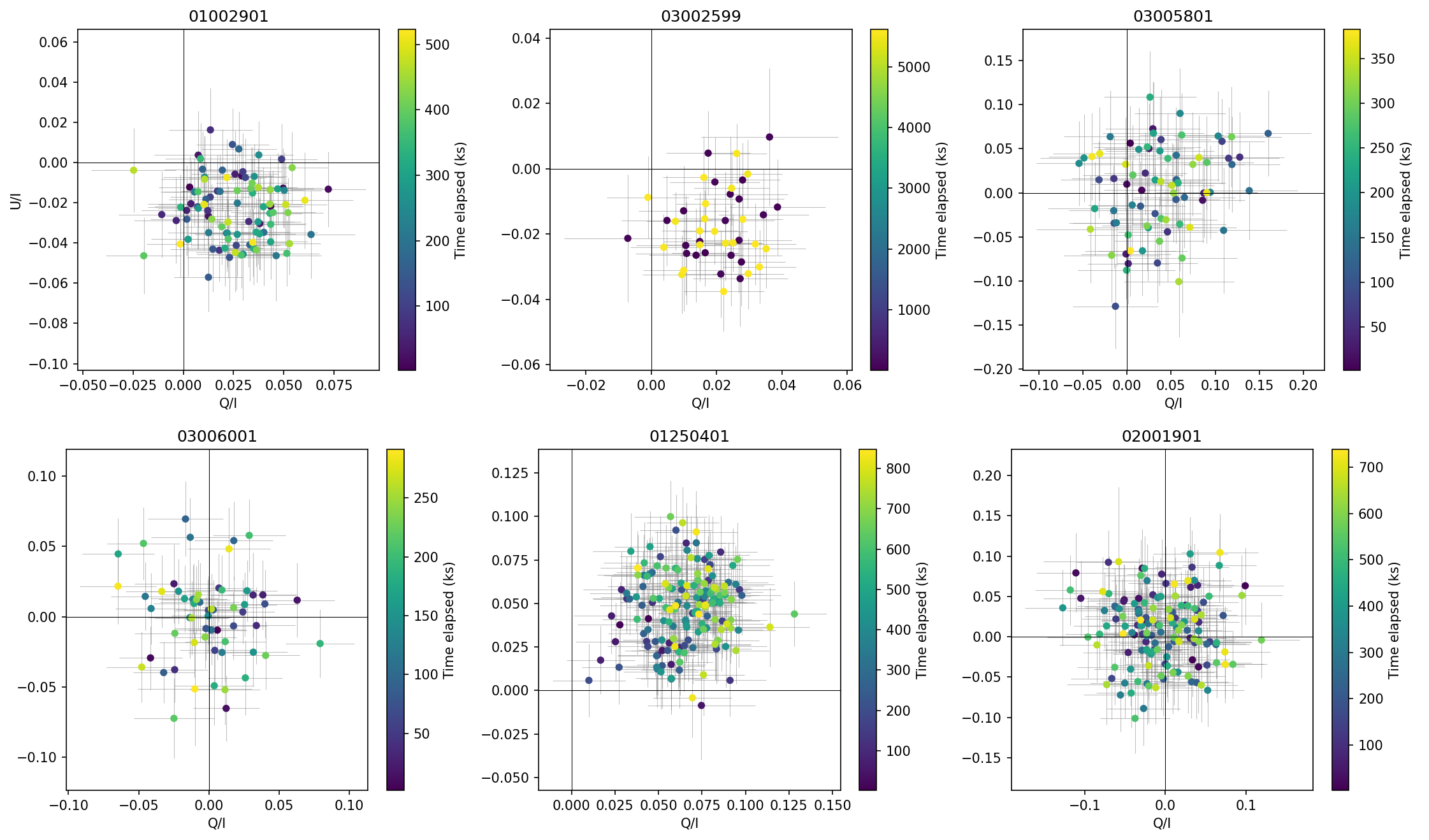}
\caption{Normalized Stokes parameters of the 5~ks segments of the six observations of the
stability analysis (panels by observation: 01002901 and 03002599, Cyg X-1; 03005801 and 03006001,
Swift J1727.8$-$1613; 01250401, 4U 1630$-$47; 02001901, LMC X-1), produced by the plotting tool written by the pipeline
(Sect.~\ref{ss:events}). Each point is one segment with its errors, coloured by elapsed time. The
constancy test is the weighted $\chi^2$ of these points about their weighted mean. The points of
observation 01250401 lie away from the origin; those of 03006001, which has no detected
polarization, scatter around it (Sect.~\ref{ss:res-stability}).}
\label{f:stokesplane}
\end{figure*}

For five of the six observations, $(q,u)$ is consistent with being constant
(Fig.~\ref{f:stokesplane}), with $p$ between 0.75 and 0.98. The exception is the Swift
J1727.8$-$1613 observation 03006001 ($\chi^2=155.9$ for 116 degrees of freedom, $p=0.0079$). For
this data set, no significant polarization was detected (PD $=(0.05\pm0.33)\%$). After accounting for the six trials, the $p$-value is 0.046, so the variability is marginal. No observation shows significant PD variability.
Fig.~\ref{f:stability} shows the PA and PD time series.

\subsection{Polarization--hardness survey}
\label{ss:res-survey}

In the third analysis (see also \PaperI), the pipeline correlated the 2--8 keV PD with the spectral
hardness for 29 IXPE observations of six BHXRBs: Cyg X-1 (12 observations), Swift J1727.8$-$1613
(8), LMC X-3 (4), 4U 1630$-$47 (2), GX 339$-$4 (2) and LMC X-1 (1). The pipeline computed the hardness ratio $\mathrm{HR}=(H-S)/(H+S)$ from the Stokes $I$ count rates in the 2--4 keV ($S$) and 4--8 keV ($H$) bands, summed over the three detector units, after subtracting an annulus background. It subtracted the background from the PD of LMC X-1, LMC X-3 and Swift J1727.8$-$1613, but not from that of the bright sources Cyg X-1 and 4U 1630$-$47 \citep{2023AJ....165..143D}. The PD exceeds the MDP99 in 24 of the 29
observations, from four sources. The five observations without a detection lie at the soft end of the HR range.

For the 24 detections, PD and HR are strongly correlated (Spearman $\rho=0.947$,
$p=2.5\times10^{-12}$, Fig.~\ref{f:pdhr}). However, the hardness ratios are not corrected for absorption. The two observations of the
heavily absorbed source 4U 1630$-$47 therefore appear harder than the hard state of Cyg X-1, and
they lie at the top right of Fig.~\ref{f:pdhr}. They contribute to the combined correlation, and
the pipeline did not recognize this problem. The correlations within single sources are thus
more meaningful. The pipeline tested each source with at least four detections. The correlation
holds within Cyg X-1 and Swift J1727.8$-$1613, and is not significant within LMC X-3 (Fig.~\ref{f:pdhrsource}). The trend agrees with
the published decrease of the PD as Swift J1727.8$-$1613 softened \citep{svoboda2024,ingram2024} and
 with the correlations within these sources in Fig.~3 of \PaperI.
 \PaperI\ analysed the same 29 observations. Its reviewer agent suspected that the combined correlation was an artifact of mixing sources.

\subsection{The $q$--$u$ time lag}
\label{ss:res-qulag}

In the fourth analysis, the pipeline searched for a time lag between the normalized Stokes
parameters $q(t)=Q/I$ and $u(t)=U/I$ within single observations. A non-zero lag would indicate a
rotation of the polarization vector. We know of no published measurement. The pipeline read
45 of the 47 requested observations. The other two are still proprietary and encrypted, and the
pipeline reported them as not measured.

 The pipeline divided each observation into time slices with enough events to reach a minimum detectable polarization at 99\% confidence (MDP99) of
6\%, the value set in the request. It assigned each slice the mean arrival time of its events. The pipeline then computed the discrete correlation function (DCF) of \citet{edelson1988} on
the resulting irregular time grid. To estimate the significance of the largest $|\mathrm{DCF}|$
over all lags, it shuffled the time order of the $u$ values 1{,}000 times, with $q$ and the time
stamps fixed. It then compared the observed maximum with the maxima of the shuffled data. This method
can underestimate the chance probability of a DCF peak, because it ignores the time coherence of the
$q$ and $u$ light curves. Spectral methods that preserve the autocorrelation of the $q$ and $u$
variations would be more reliable (see \citealp{timmer1995}, \citealp{uttley2014}, and references
therein).

\begin{figure*}
\centering
\includegraphics[width=0.95\textwidth]{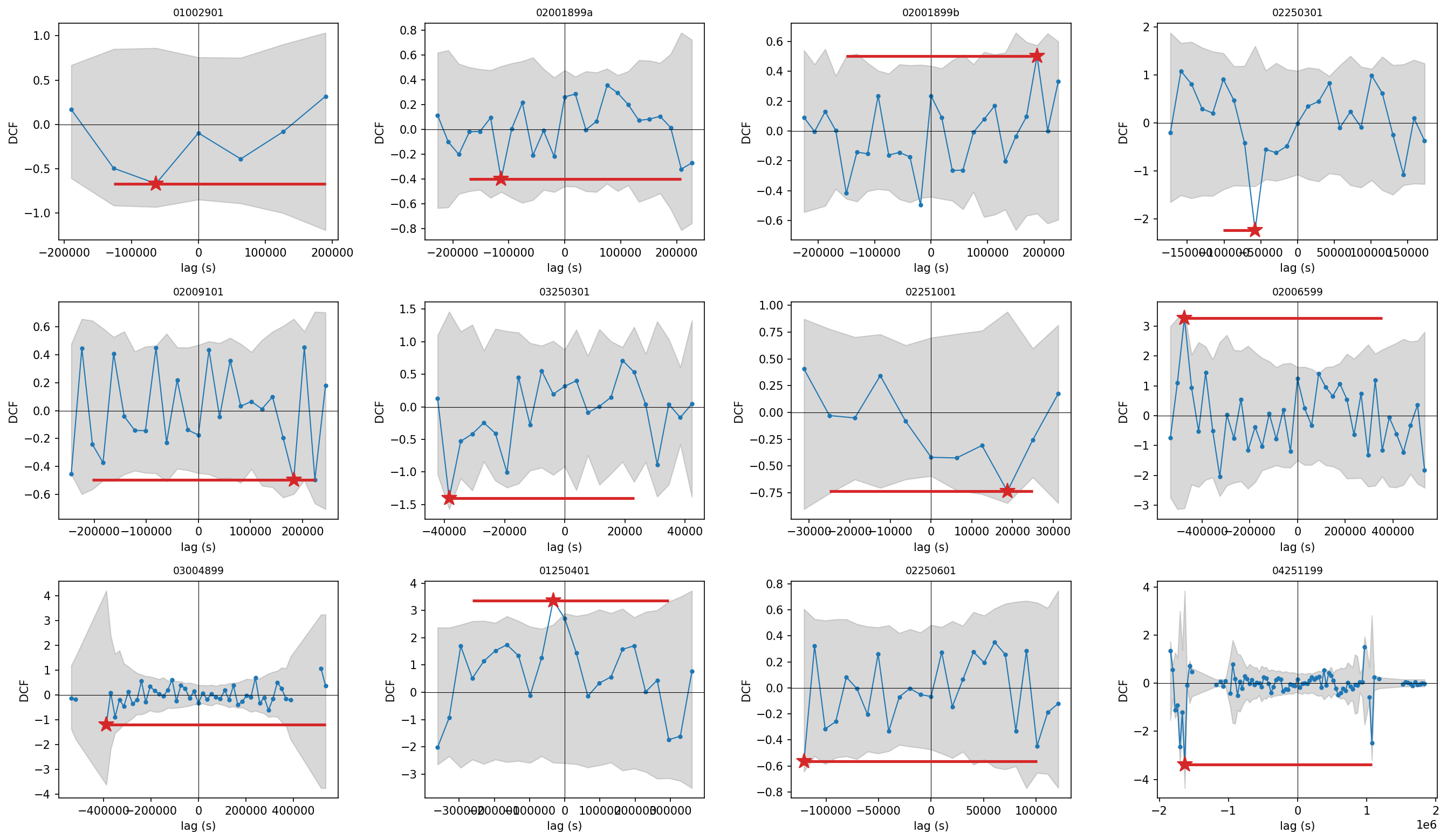}
\caption{Lagged $q$--$u$ discrete correlation function for the 12 segments of the time-lag search, produced by the tool \texttt{ixpe\us{}qu\us{}lag\us{}dcf\us{}plot}. Grey band: 1st to 99th percentile of 1{,}000 time-shuffled permutations. Star: the peak. Red bar: the 68\% flux-randomization interval of the peak lag. Cyg~X-3 observation 02250301 (top right) shows the only candidate.}
\label{f:qulag}
\end{figure*}

 Eleven of the 45 observations have enough events for at least eight slices, and the pipeline
excluded the other 34. The 11 observations yield 12 segments from six sources, because one
observation was split into two. Fig.~\ref{f:qulag} shows the DCFs. Eleven of the 12 segments show no significant lagged correlation. Cyg X-3 observation 02250301 has a DCF peak at a lag of $-57.7$~ks with $p=0.0060$, below the threshold of 0.01. After accounting for the 12 segments tested, $p=0.07$ \citep{sidak1967}, so the pipeline reported this lag as a candidate, not a detection. \ref{a:qulag} describes the tool writing for this analysis.

\begin{table*}
\centering
\caption{Integrity problems in the tool-writing sessions of the analyses (Sect.~\ref{s:campaign})
and of the trials of Sects.~\ref{ss:trials} and~\ref{s:generality}, grouped by class. ``Caught
by'' names the agent or the gate check of Sect.~\ref{ss:selfext}.}
\label{t:events}
\small
\begin{tabular}{@{}L{2.1cm}L{5.2cm}L{2.7cm}L{4.4cm}@{}}
\hline
Class & Problem & Caught by & Outcome \\
\hline
Fabrication
 & Test files hard-coding science results, while claiming they come from real data & co-lead & Rejected and reworked \\
\cline{2-4}
 & Documentation claiming test items that do not exist & co-lead & Rejected and reworked \\
\cline{2-4}
 & Synthetic data for a test of the tool's capability did not exist & coder $\rightarrow$ lead escalation & The lead declined to generate data; the committed test writes hand-picked tables (\ref{a:toolspec}) \\
\hline
Defect in code written by the pipeline
 & Plotter silently failing when inputs missing & co-lead & Rejected, reworked, and fixed \\
\cline{2-4}
 & Tests not fault-detecting & lead $\rightarrow$ coder; co-lead & Rejected and fixed \\
\cline{2-4}
 & Accepting the null hypothesis where significance was undefined & co-lead & Rejected and fixed; unresolved in one trial \\
\cline{2-4}
 & A spectral fit operating on every loaded data group while documented as single-mission & co-lead & Rejected before commit \\
\hline
Defect in trusted code
 & Indexing error in a human-written reduction helper that dropped IXPE detector unit 3, and uncertainties added instead of variances & reported by a human & Fixed by the coder, with a regression test \\
\hline
Failure of a check
 & False rejections on banned vocabulary & gate: banned vocabulary & Reworded \\
\hline
\end{tabular}
\end{table*}

\section{Benchmarking the tool-writing pipeline}
\label{s:verification}

 In this section, we test the tool-writing pipeline (Fig.~\ref{f:authoring}). Section~\ref{s:generality} tests the analysis pipeline (Fig.~\ref{f:arch}). We first list the
integrity problems that occurred during tool writing (Sect.~\ref{ss:events}). We then plant defects in tools and count how many the checks stop. We repeat the count with one check disabled at a time (Sect.~\ref{ss:catchrate}). Finally, we measure the reliability and cost of writing the same tool repeatedly with four different coders (Sect.~\ref{ss:trials}).

\subsection{Integrity problems in the tool-writing sessions}
\label{ss:events}

 Table~\ref{t:events} lists the problems that occurred during tool writing and how each was
detected. As in \citet{sciintegrity2026}, most problems arose under pressure to finish, here a nearly used-up submission budget, or when input data were missing. For example, the synthetic data that one test required did not exist,
and the coder asked the lead how to proceed (\ref{a:transcripts}).

 The checks themselves can be wrong. The most frequent error was a false rejection by the banned-vocabulary scan. We accept these false rejections because they are cheap. A false rejection uses one submission, and the coder rewords the text and submits again. In the trials of Sect.~\ref{ss:trials}, the gate rejected fewer submissions than the co-lead, and every failed trial ended with an objection of the co-lead still open (Table~\ref{t:sweepevents}).

\subsection{Catch rate of planted defects}
\label{ss:catchrate}

\begin{figure*}
\centering
\includegraphics[width=\textwidth]{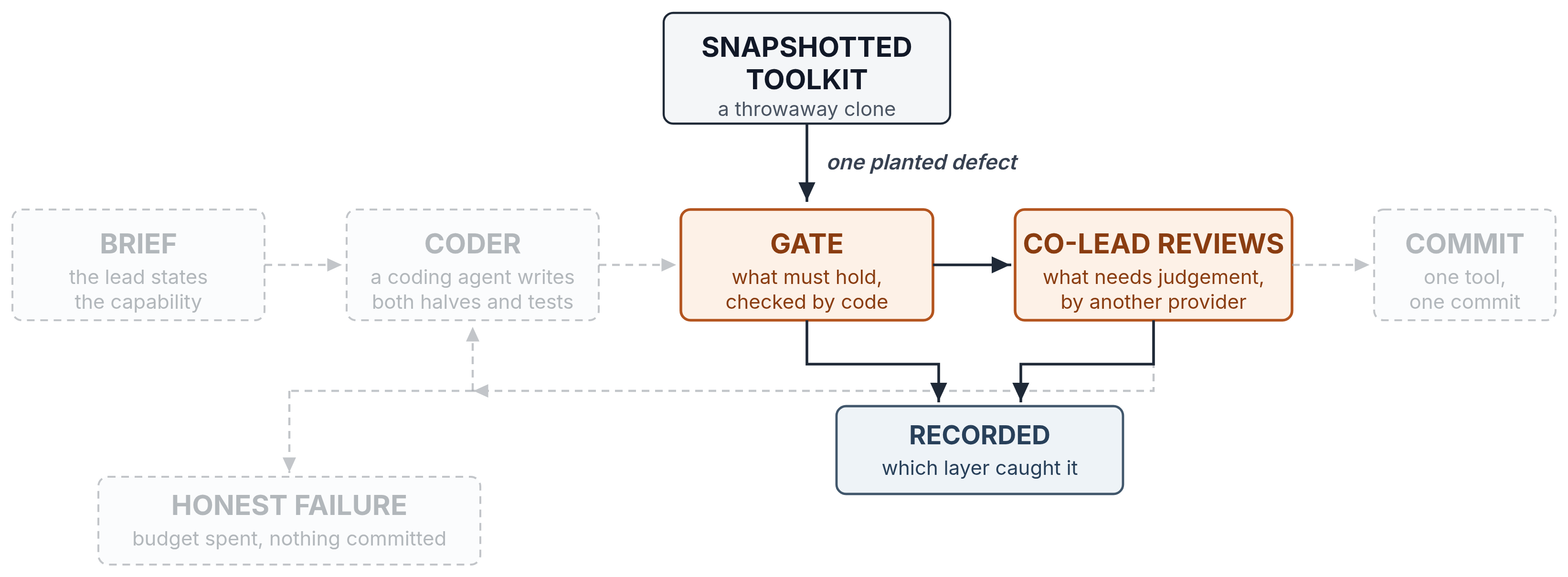}
\caption{Path of a planted defect through the checks. The benchmark inserts the defect into a copy of the toolkit and passes it to the gate in place of the output of a tool-writing session.}
\label{f:benchmark}
\end{figure*}

We know of no established benchmark for how well checks catch defects in AI-written scientific
code, so we built one. The benchmark inserts each defect into a copy of the toolkit and passes the modified toolkit through the gate and the code review (Fig.~\ref{f:benchmark}). It contains 65 planted defects and 69 defect-free control changes (Table~\ref{t:catchrate}).

The defects fall into six classes. \emph{Fabrications} invent values, with or without a banned word in the
text. \emph{Wrong results} include a PA sign flip, a degree/radian confusion, an inverted calibration
epoch, bin edges shifted by half a bin, and 11 errors taken from objections of the co-lead during our analyses. We plant each of the first four twice, once with a test that exposes the error and once with a test that misses it. \emph{Tests that check nothing} have no assertions, cannot fail, or are
skipped. \emph{Hidden limits} are a plausible fixed constant or an upstream product accepted as complete.
\emph{Missing structure} is a missing test file or manifest entry. Six further defects are \emph{invisible to the gate}, so only the co-lead can stop them.

Table~\ref{t:catchrate} gives the results for the default configuration. The checks stopped 64 of the 65 planted defects. All controls passed the gate, but the co-lead rejected 40 of the 69, mostly because it judged their tests incomplete (see below). An LLM read the stated reason for each rejection of a defect and checked whether it named the planted defect. Of the 46 defects that the co-lead rejected, 44 were rejected for the planted defect. In the eight-model comparison below, 295 of 308 rejections named the defect in the specific code.

\begin{figure*}
\centering
\input{tables/catchrate.tex}

\vspace{1.2em}
\centering
\includegraphics[width=\textwidth]{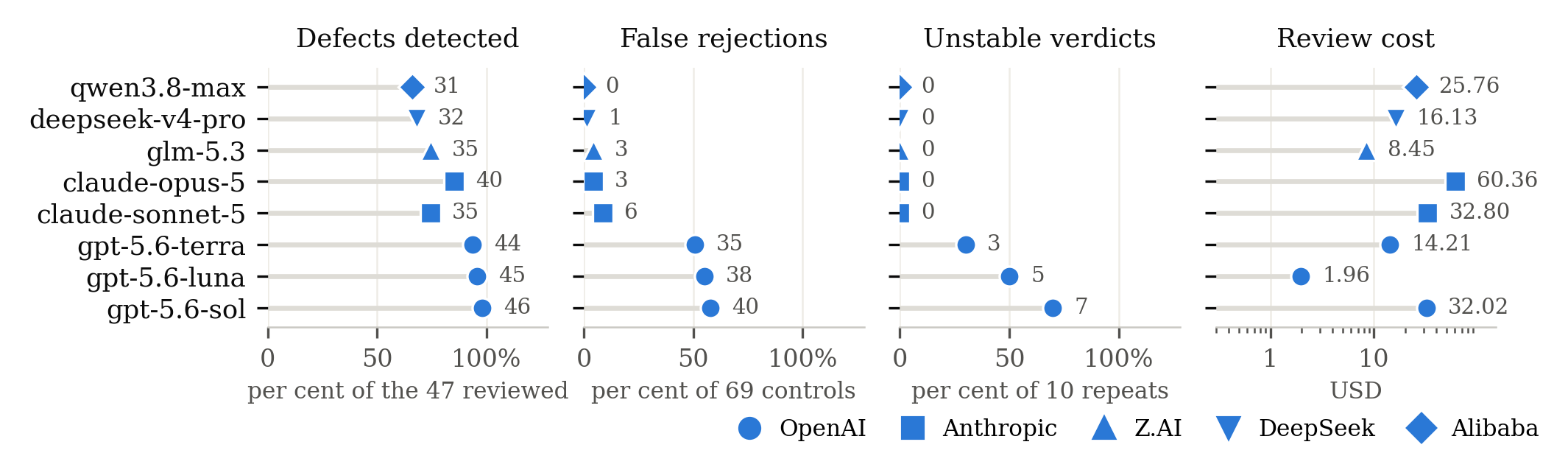}
\captionof{figure}{Code review by eight models from five providers, on the defects that pass the
gate and on the 69 controls of Sect.~\ref{ss:catchrate}. Unstable verdicts are counted on 10 repeated submissions of the same defect-free package. A rejection of one of these submissions contradicts the model's acceptance of the same input.}
\label{f:modelswap}
\end{figure*}

\label{ss:ablations}
 We re-ran the benchmark with one check disabled at a time (Table~\ref{t:ablations}). In each run with a gate check disabled, the co-lead caught every defect that this check stopped in the default configuration. Without the co-lead review, most defects reached a commit.

\label{ss:modelswap}
We also compared eight models from five providers as the co-lead (Fig.~\ref{f:modelswap}). The
 models detect between 31 and 46 of the 47 defects that reach the review and falsely reject between 0 and 40 of the 69 controls. The three GPT models detect the most defects and also falsely reject the most controls. In all 113 false rejections, the GPT models judge the tests inadequate, and some also object to the code. The controls contain no planted defect, but most of these objections name real weaknesses in their tests or code. The GPT verdicts also vary between repeated submissions of the same defect-free package (Fig.~\ref{f:modelswap}). When a provider returned no verdict, we counted a defect as not detected and a control as not falsely rejected.\footnote{Six reviews returned no usable verdict: three by \texttt{deepseek-v4-pro}, two by \texttt{qwen3.8-max} and one by \texttt{glm-5.3}. We request each verdict as a forced tool call or a structured output. DeepSeek states that its JSON output may occasionally be empty (\url{https://api-docs.deepseek.com/guides/json_mode}). Alibaba does not support a forced tool call in thinking mode, which our runs used (\url{https://www.alibabacloud.com/help/en/model-studio/qwen-function-calling}), and Z.AI supports none (\url{https://docs.z.ai/guides/capabilities/function-calling}).}

The choice of co-lead is a trade-off between detection and false rejection. Each false rejection
costs one of the five submissions and one more review. We chose \texttt{gpt-5.6-sol} because it detects the most defects, at the cost of the most false rejections (Fig.~\ref{f:modelswap}).

\begin{table}
\centering
\caption{Defects reaching a commit with one check disabled at a time, on the 65 defects of
Table~\ref{t:catchrate}. In both runs with a gate check disabled, the co-lead rejected the defect that had reached a commit in the default configuration, so its verdict on this unchanged package differed between runs (Fig.~\ref{f:modelswap}).}
\label{t:ablations}
\small
\begin{tabular}{@{}L{4.4cm}c@{}}
\hline
Configuration & Reaching \\
 & a commit \\
\hline
Default & 1/65 \\
No scan & 0/65 \\
No tool-local tests & 0/65 \\
No co-lead review & 46/65 \\
\hline
\end{tabular}
\end{table}

\subsection{Reliability and cost of tool writing}
\label{ss:trials}

\begin{figure*}
\centering
\captionof{table}{ Tool-writing trials of Fig.~\ref{f:trialcost}, with the lead (\texttt{claude-opus-5}) and the co-lead held fixed. $N$ is the number of trials, and rejections are per trial.}
\label{t:sweepevents}
\small
\begin{tabular}{@{}L{3.6cm}L{4.2cm}cccccc@{}}
\hline
Coder & Co-lead & $N$ & Committed & Gate & Co-lead & Cost US\$ & Minutes \\
 & & & & rej./trial & rej./trial & (median) & (median) \\
\hline
deepseek-v4-flash & gpt-5.6-sol & 5 & 5 & 0.6 & 1.6 & 6.5 & 52 \\
glm-5.3 & gpt-5.6-sol & 5 & 4 & 0.2 & 2.8 & 16.0 & 90 \\
claude-sonnet-5 & gpt-5.6-sol & 10 & 7 & 1.4 & 2.2 & 82.0 & 116 \\
gpt-5.6-terra & gpt-5.6-sol \emph{(same provider)} & 5 & 5 & 0.0 & 2.8 & 10.8 & 43 \\
\hline
\end{tabular}

\vspace{1.2em}
\centering
\includegraphics[width=0.8\textwidth]{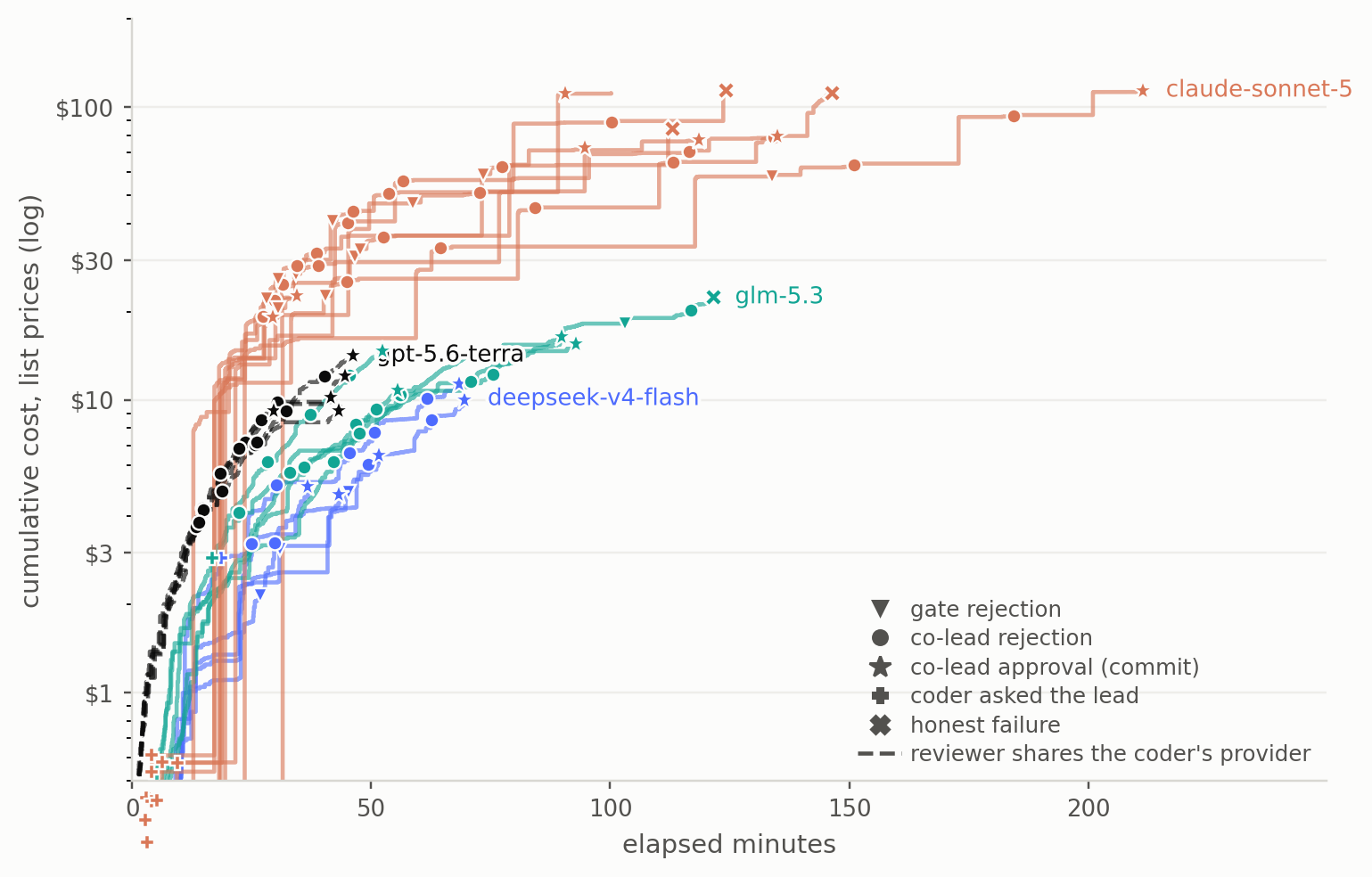}
\captionof{figure}{ Cumulative cost of each tool-writing trial against elapsed time, at API list prices, summed over the coder, the lead and the co-lead. The cost axis is logarithmic. Colour gives the provider, shape the event
($\blacktriangledown$ gate rejection, $\bullet$ co-lead rejection, $\star$ co-lead approval and
commit, $\mathbf{+}$ coder question to the lead, $\times$ declared failure). A dashed line marks the \texttt{gpt-5.6-terra} trials, whose co-lead shares the coder's provider.}
\label{f:trialcost}
\end{figure*}

 To measure the reliability and cost of tool writing, we had the pipeline write the same tool in repeated trials, following \citet{atlas2025}. Each trial started from a fresh copy of the toolkit.
Every trial used the brief that an earlier analyst lead (\texttt{claude-opus-4-8}) wrote during the
stability analysis (\ref{a:toolspec}). No human took part. In all trials, we fixed the lead (\texttt{claude-opus-5}) and the co-lead (\texttt{gpt-5.6-sol}), varying the coder. We ran 10 trials with \texttt{claude-sonnet-5} as the coder and five with each other coder. Table~\ref{t:sweepevents} summarizes the trials, and Fig.~\ref{f:trialcost} shows their cost over time. With \texttt{claude-sonnet-5}, the pipeline committed a tool in 7 of the 10 trials. Each of the three failed trials ran out of submissions with an objection of the co-lead still open.

\begin{sloppypar}
With \texttt{claude-sonnet-5}, the coder accounted for a median of 89\% of the cost of a trial. With \texttt{deepseek-v4-flash}, all five trials committed a tool, at a median total cost of US\$6.5 per trial and a median coder cost of US\$0.68. The median total cost with \texttt{claude-sonnet-5} was about 13 times higher. With so few trials, the difference in commit rate (5 of 5 against 7 of 10) is consistent with chance. The coders also made changes of different scope. No \texttt{deepseek-v4-flash} commit changed the reduction module of the toolkit, whereas nearly all commits by the other coders did, so its higher commit rate may reflect a smaller change.
\end{sloppypar}

The tool-writing benchmark has logged 252 coder sessions since it began, including the 25 trials of Table~\ref{t:sweepevents}. Over these sessions the gate rejected 68 of 386 submissions, mostly for banned vocabulary (Table~\ref{t:gateevents}). In the 25 trials of
Table~\ref{t:sweepevents}, banned vocabulary caused all 18 gate rejections. Some of these rejections were false positives (Sect.~\ref{ss:events}).

\begin{table}
\centering
\caption{Gate rejections by failing check, over the 386 gate runs of all 252 logged coder sessions of the tool-writing benchmark.
  A rejected submission can fail more than one check.}
\label{t:gateevents}
\small
\begin{tabular}{@{}lrr@{}}
\hline
Failing check & Rejections & Of gate runs \\
\hline
Banned vocabulary & 52 & 13.5\% \\
Tool-local tests & 10 & 2.6\% \\
Structure & 6 & 1.6\% \\
Determinism & 1 & 0.3\% \\
Parseability & 1 & 0.3\% \\
\hline
All rejections & 68 & 17.6\% \\
Gate runs & 386 & \\
\hline
\end{tabular}
\end{table}

\input{tables/seats.tex}
\input{tables/slots.tex}

\begin{figure*}
\centering
\includegraphics[width=\textwidth]{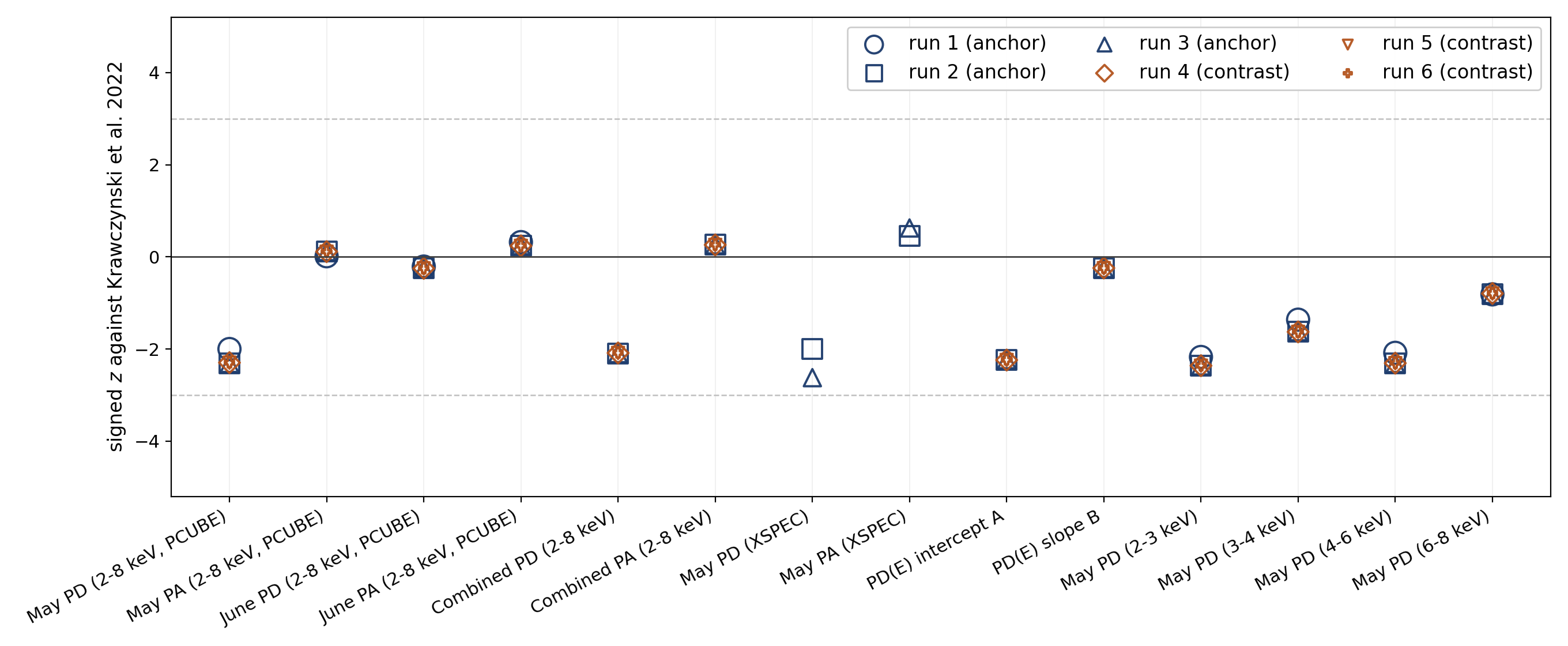}
\caption{Signed $z$ of the 14 prescribed quantities in the six runs of the model-tier test. Hollow markers show the anchor assignment, filled markers the contrast assignment, and dashed lines $|z|=3$. Each column is one quantity. The reports of the six runs are provided
as ancillary files (\texttt{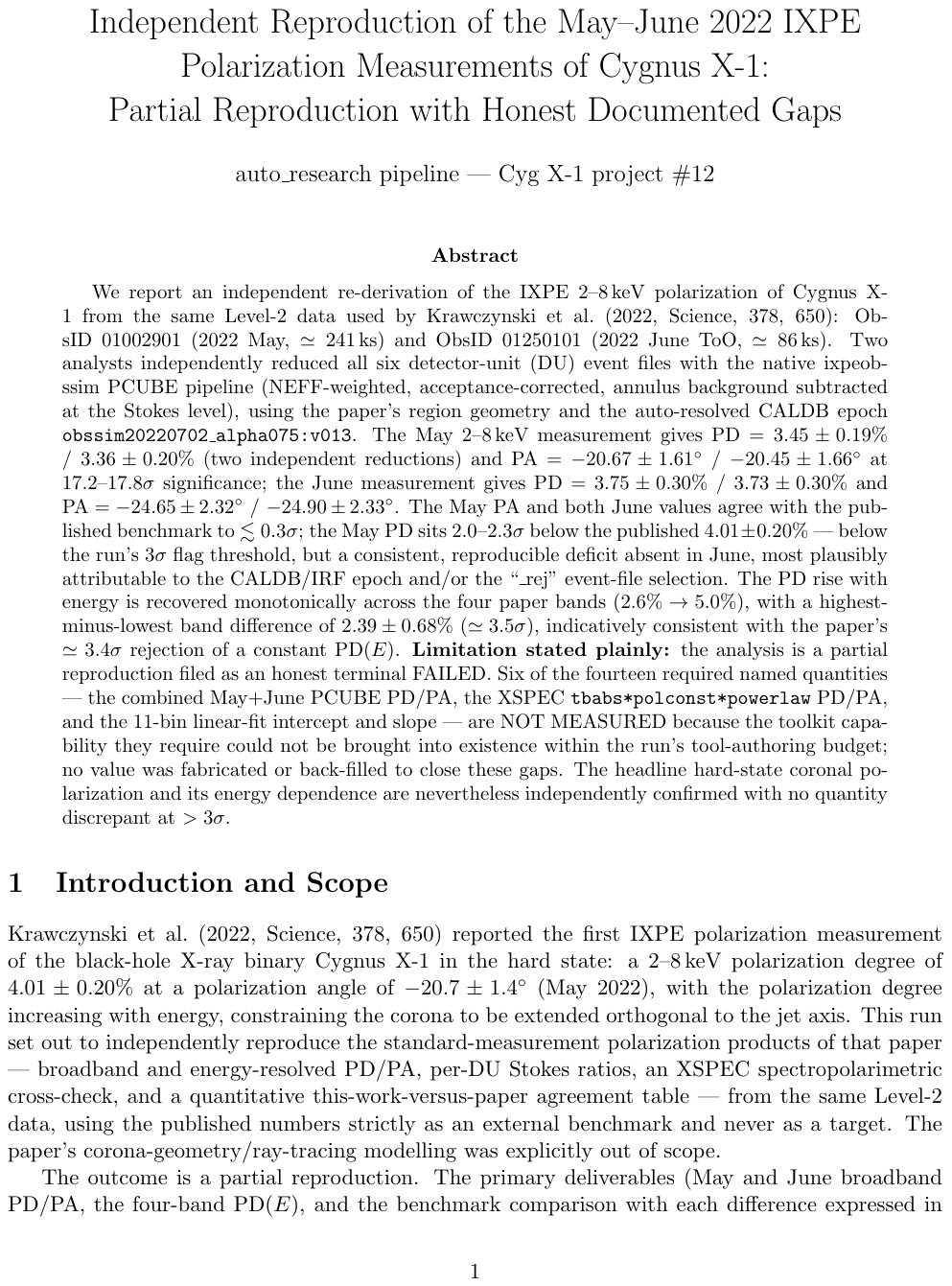} to \texttt{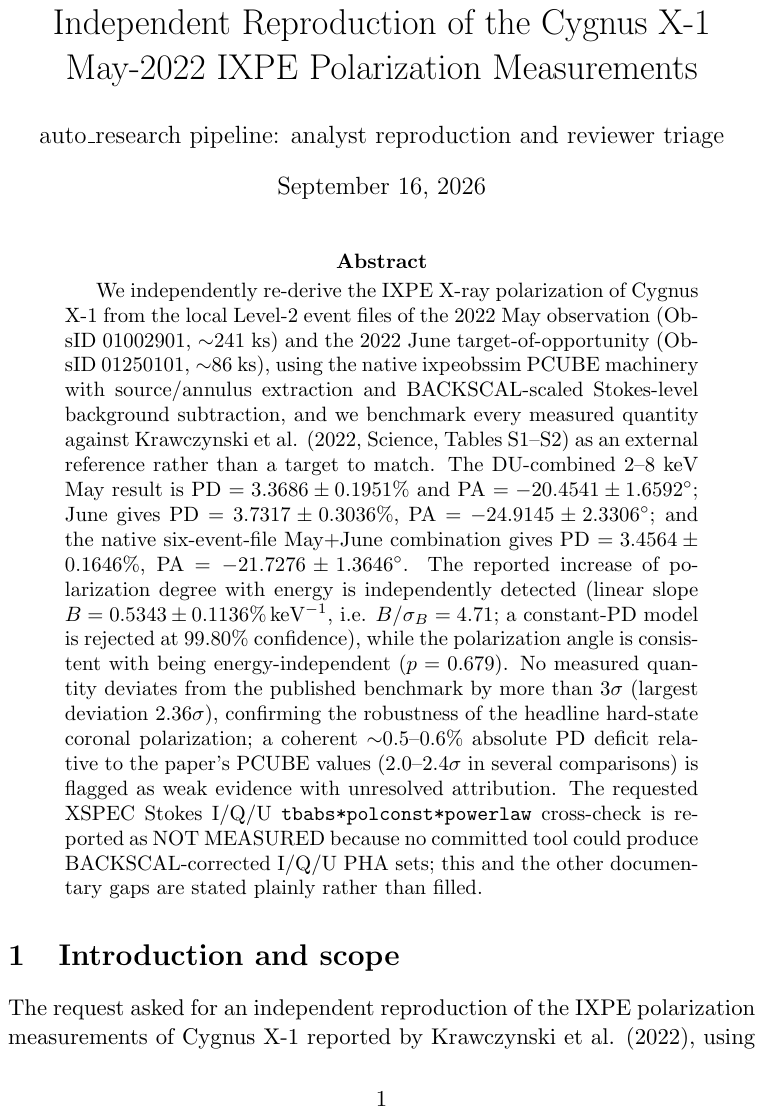}) with the arXiv version of this paper.}
\label{f:overlay}
\end{figure*}

\section{Generality across model tiers}
\label{s:generality}
\label{ss:tiers}

We test whether the analysis pipeline still produces a verifiable result, or reports that it
cannot, when weaker models fill its main roles. We keep the coder (\texttt{deepseek-v4-flash}) and
the co-lead (\texttt{gpt-5.6-sol}) of Sect.~\ref{s:verification}, and change the models of the
analyst lead, the reviewer and one of the two analyst workers together. The default models form the
``anchor'' assignment. Models that score lower on published benchmarks for the tasks of each role
form the ``contrast'' assignment (Table~\ref{t:seats}). We ran the Cyg X-1 request of
Sect.~\ref{ss:res-calibration} three times with each assignment, with the same brief, toolkit
snapshot and observations, and with tool writing enabled.

 Three runs per assignment are too few to measure run-to-run consistency \citep{stochasticity2512}. We therefore describe how each run handled the same task, which required tools missing from the toolkit. Two runs wrote these tools and recovered all 14 quantities. The other four reported the rest as not measured. Each report thus gives a verifiable value or states that the value is missing.

Table~\ref{t:slots} lists the six runs, and Fig.~\ref{f:overlay} compares the recovered
quantities. All runs reproduced the systematic PD offset of Sect.~\ref{ss:res-calibration}. All runs started from the same toolkit snapshot and data, and five of them wrote their own tools on top of it. The two complete anchor runs and the three contrast runs agree closely on the 12 quantities that all five recovered (Fig.~\ref{f:overlay}).

In the first anchor run, the coding session ended in a declared failure. The co-lead rejected all five submissions. The lead then reported that the required capability was unavailable, and the reviewer accepted this report as a partial reproduction. The run recovered eight of the 14
quantities with the tools already in the toolkit (Table~\ref{t:slots}). All rejections concerned the
robustness of the code, for example the reuse of stale output files and exposure keywords taken from
the first input only. The number of objections per rejection fell from 4 in the first to 2 in the last.

None of the contrast runs produced a working XSPEC \texttt{polconst} tool, and the toolkit held no
substitute. In run~5, one analyst worker computed the May PD and PA of this fit with PyXspec in a shell session during the first round. The reviewer later told the analysts to record the two quantities as not measured if no toolkit tool could fit the model, and to write no further tools in the next round.
The values reached the lead in the worker's first-round report. The lead did not carry the values of the shell fit forward, because the fit lacked background subtraction. This exclusion agrees with the rule of Sect.~\ref{ss:report} that shell output supports no number. The final report records both quantities as not measured.

\section{Discussion}
\label{s:limitations}
\label{ss:notyet}
\label{ss:otherlimits}
\label{ss:revision}

Unlike systems that write new code for each task
\citep{2025arXiv251026887V,cmbagent2025,Lu2024,aicosmologist2025}, \pipeline\ keeps the tools that
it writes. A tool passes the checks once and is then available to later analyses. A human can
check it by running its library functions directly. The gate, the report validator, the two-repository boundary and the rule for promoting tools contain no code specific to an instrument or to astronomy. They could therefore serve other quantitative analyses.

Our results are based on four analyses of similar data. We have not yet measured how often later
runs revise the conclusions of earlier ones, how often later analyses reuse committed tools, or how
much it costs a human to verify a tool. The design has three limitations. First, provenance enforcement establishes the origin of a number, not its correctness. A defective tool that passes the gate and the review yields numbers that the validator admits. Second, the co-lead is a language model and therefore fallible. In the benchmark of Sect.~\ref{ss:catchrate}, it let one of the 47 defects that passed the gate reach a commit and rejected 40 of the 69 defect-free controls. Third, the validator verifies that a number appears in a cited output, not that it was taken from the correct place in that output.

\section{Summary and outlook}
\label{s:outlook}

We have presented \pipeline, a semi-autonomous multi-agent pipeline that analyses X-ray
polarimetry data and writes the analysis tools that it needs. It checks its work at two points. A
new tool passes automatic checks and a review by a model from a different provider before an analysis uses it, and a human decides whether to promote it to the trusted branch. The validator accepts a number in a report only if a cited output of a toolkit tool or the request
contains it, or if the report gives its location in a cited paper.

In four analyses of IXPE data, the pipeline wrote 13 tools and extended four. It reproduced the Cyg X-1 results of
\citet{2022Sci...378..650K} with all 14 quantities within $3\sigma$ and found a systematic offset of the PD values. The checks stopped 64 of the 65 planted defects. With weaker models in the main roles, the pipeline recovered 12 of the 14 quantities. It reported the other two as not measured, because no toolkit tool could fit the required model, and stated no value for them, even in a run in which an agent had computed them in a shell. In contrast, the models of \citet{sciintegrity2026} produced synthetic data in 55.6\% of runs with insufficient data, even when they could report a failure.

In \pipeline, the reviewer, the librarians and the analysts each work with their own tools and can end a task only through a report that the validator checks. We plan to build these agents, and new ones, from one reusable agent design. A team of any structure could then be assembled from such agents, and an agent could start new agents during a run, so that the team adapts to the task while the validator still checks every report. The code of the pipeline and of the toolkit is public (see Data availability).

\section*{CRediT authorship contribution statement}
\textbf{Fang Zhou:} Methodology, Software, Investigation, Writing -- original draft,
Visualization.

\section*{Declaration of competing interest}
The author declares that they have no known competing financial interests or personal
relationships that could have appeared to influence the work reported in this paper.

\section*{Acknowledgments}
The author thanks Kun Hu for discussions and comments on the manuscript, and Henric Krawczynski and the members of the AI group of the Physics
Department at Washington University in St. Louis for joint discussions and the sharing of resources
(research project ideas, downloaded data). This research used data and software provided by the
High Energy Astrophysics Science Archive Research Center (HEASARC), a service of the Astrophysics
Science Division at NASA/GSFC. The author acknowledges support from the Physics Department at
Washington University in St. Louis through its Hufferd Seed Funding program.

\section*{Declaration of generative AI and AI-assisted technologies in the manuscript
preparation process}
 The pipeline is built from LLM agents (\ref{a:impl}). During the preparation of this work the author used Claude (Anthropic) in order to write the pipeline's code and working notes for the manuscript. The author reorganised these notes into the first draft and revised it over several rounds. After using this tool, the author reviewed and edited the content as needed and takes full responsibility for the content of the published article.

\section*{Data availability}
The IXPE data are public at HEASARC. The code of the pipeline and of the toolkit, including the
tools that the pipeline wrote, is available at \url{https://github.com/tomthestuka/auto_research}
and \url{https://github.com/tomthestuka/research_toolkit}. The event logs of the runs and the
planted defects and controls of Sect.~\ref{ss:catchrate}, with the review verdicts, are available
from the corresponding author on request. The reviewer's reports of the six model-tier runs of
Sect.~\ref{s:generality} are provided as ancillary files with the arXiv version.

\bibliographystyle{elsarticle-num-names}
\bibliography{FANGZ}

\appendix
\counterwithin{figure}{section}
\counterwithin{table}{section}
\renewcommand{\thefigure}{\Alph{section}.\arabic{figure}}
\renewcommand{\thetable}{\Alph{section}.\arabic{table}}

\input{appendix.tex}

\end{document}

%% file: tables/campaign.tex
\begin{table*}
\centering
\caption{Summary of the analyses. ``Tools'' is the number of tools committed for the analysis.
For the Cyg X-1 reproduction, one of the seven extends the data loader; for the time-lag search,
all three extend existing tools. ``Cost'' is the cost of the model calls at list prices. ``Time''
is the running time of the analysis.}
\label{t:campaign}
\small
\begin{tabular}{@{}L{4.3cm}lclrL{3.4cm}@{}}
\hline
Analysis & Data & Tools & Cost (US\$) & Time & External check \\
\hline
Cyg X-1 polarization reproduction (Sect.~\ref{ss:res-calibration}) & 2 ObsIDs & 7 & 55 & 7.0~h & published values \citep{2022Sci...378..650K} \\
Polarization stability catalogue (Sect.~\ref{ss:res-stability}) & 6 ObsIDs & 3 & 35 & 2.4~h & internal, literature \\
Polarization--hardness survey (Sect.~\ref{ss:res-survey}) & 29 ObsIDs & 4 & 45 & --- & \PaperI, literature \\
$q$--$u$ time-lag search (Sect.~\ref{ss:res-qulag}) & 45 ObsIDs & 3 & 377 & 12.0~h & none \\
\hline
\end{tabular}
\end{table*}

%% file: tables/cygx1.tex
\begin{table*}
\centering
\caption{Reproduction of the Cyg X-1 measurements of \citet{2022Sci...378..650K}. The signed
deviation is $z=(\mathrm{pipeline}-\mathrm{published})/\sqrt{\sigma_{\rm pipeline}^2+\sigma_{\rm published}^2}$,
computed from unrounded values. The pipeline values come from its tool
\texttt{compare\us{}to\us{}benchmark}; the published values of the four energy bands are those of
Table S2 of the reference (``PD - sum''). The pipeline's XSPEC rows fit
\texttt{tbabs*polconst*powerlaw}; the reference does not name the model of its XSPEC values. The
pipeline's intercept $A$ and slope $B$ come from an XSPEC fit of \texttt{tbabs*pollin*powerlaw},
with PD$(E) = A + B\,(E/\mathrm{keV}-1)$; the published $A$ and $B$ come from a linear fit of the
same form to PCUBE values.}
\label{t:cygx1}
\small
\begin{tabular}{llccr}
\hline
Quantity & Unit & Pipeline & \citet{2022Sci...378..650K} & $z$ \\
\hline
May PD (2--8 keV, PCUBE) & \% & $3.37 \pm 0.20$ & $4.01 \pm 0.20$ & $-2.29$ \\
May PA (2--8 keV, PCUBE) & deg & $-20.5 \pm 1.7$ & $-20.7 \pm 1.4$ & $+0.11$ \\
June PD (2--8 keV, PCUBE) & \% & $3.73 \pm 0.30$ & $3.84 \pm 0.31$ & $-0.25$ \\
June PA (2--8 keV, PCUBE) & deg & $-24.9 \pm 2.3$ & $-25.7 \pm 2.3$ & $+0.24$ \\
Combined PD (2--8 keV) & \% & $3.46 \pm 0.17$ & $3.95 \pm 0.17$ & $-2.09$ \\
Combined PA (2--8 keV) & deg & $-21.7 \pm 1.4$ & $-22.2 \pm 1.2$ & $+0.26$ \\
May PD (XSPEC) & \% & $3.36 \pm 0.14$ & $3.9 \pm 0.2$ & $-2.21$ \\
May PA (XSPEC) & deg & $-20.3 \pm 1.2$ & $-21 \pm 1$ & $+0.45$ \\
PD($E$) intercept $A$ (1 keV) & \% & $1.72 \pm 0.34$ & $2.9 \pm 0.4$ & $-2.24$ \\
PD($E$) slope $B$ & \% & $0.53 \pm 0.11$ & $0.58 \pm 0.15$ & $-0.24$ \\
May PD (2--3 keV) & \% & $2.58 \pm 0.25$ & $3.5 \pm 0.3$ & $-2.36$ \\
May PD (3--4 keV) & \% & $2.87 \pm 0.24$ & $3.5 \pm 0.3$ & $-1.63$ \\
May PD (4--6 keV) & \% & $3.76 \pm 0.28$ & $4.7 \pm 0.3$ & $-2.31$ \\
May PD (6--8 keV) & \% & $4.99 \pm 0.64$ & $5.8 \pm 0.8$ & $-0.80$ \\
\hline
\end{tabular}
\end{table*}

%% file: tables/catchrate.tex
\centering
\captionof{table}{Planted defects stopped by each check, by defect class, with
\texttt{gpt-5.6-sol} as the co-lead. The ``scan'' column covers gate checks 3--5 of
Sect.~\ref{ss:selfext}, and the ``tests, structure'' column covers checks 1, 2, 6 and 7. The
co-lead column counts rejections among the changes that passed the gate. Tests that check
nothing have no assertions, cannot fail, or are skipped. A hidden limit is a plausible fixed
constant or an upstream product accepted as complete. Missing structure is a missing test file
or manifest entry. Defects invisible to the gate exploit known weaknesses of the gate
checks. The controls
are defect-free changes, so each co-lead rejection of a control is a false rejection.
Fig.~\ref{f:modelswap} compares eight models as the co-lead.}
\label{t:catchrate}
\small
\begin{tabular}{@{}lccccc@{}}
\hline
Defect class & Defects & \multicolumn{2}{c}{Gate} & Co-lead & Reaching \\
 & & scan & tests, structure & & a commit \\
\hline
Fabrication, with a banned word & 8 & 6 & 1 & 1/1 & 0 \\
Fabrication, without a banned word & 6 & 0 & 1 & 5/5 & 0 \\
Wrong result, test exposes it & 4 & 0 & 4 & --- & 0 \\
Wrong result, test misses it & 15 & 0 & 0 & 15/15 & 0 \\
Tests that check nothing & 9 & 0 & 2 & 6/7 & 1 \\
Hidden limit & 10 & 0 & 0 & 10/10 & 0 \\
Missing structure or manifest & 7 & 0 & 4 & 3/3 & 0 \\
Invisible to the gate & 6 & 0 & 0 & 6/6 & 0 \\
\hline
All planted defects & 65 & 6 & 12 & 46/47 & 1 \\
Control (defect-free change) & 69 & 0 & 0 & 40/69 & 29 \\
\hline
\end{tabular}

%% file: tables/seats.tex
\begin{table*}
\centering
\caption{The three roles varied in the model-tier test (Sect.~\ref{s:generality}). For
each role, the models are compared on published benchmark scores for the abilities that the role
requires. ``Anchor'' is the default assignment. The coder and the co-lead are the same in all runs
(coder deepseek-v4-flash and co-lead gpt-5.6-sol). The scores were read on 2026-09-08, when the contrast assignment was chosen, from Artificial Analysis
(Intelligence Index and CritPt), the Terminal-Bench 4.0 leaderboard and the LMArena vision
leaderboard. $^{*}$AutomationBench-AA was read on 2026-09-22 from Artificial Analysis.}
\label{t:seats}
\small
\renewcommand{\arraystretch}{1.25}
\begin{tabular}{L{1.4cm}L{2.5cm}L{2.9cm}L{2.1cm}L{2.1cm}L{4.0cm}}
\hline
Role & Abilities & Scores consulted & Anchor & Contrast & Anchor $\rightarrow$ contrast \\
\hline
Analyst lead & planning \newline adjudication over long agent loops & AA composite \newline Terminal-Bench 4.0 & claude-opus-5 & gpt-5.6-terra & AA: 51 $\rightarrow$ 42 \newline TB4: 51.8\% $\rightarrow$ 21.5\% \\
\hline
Reviewer & scientific critique \newline figure reading \newline general capability & AA composite \newline CritPt \newline LMArena vision & claude-fable-5 & qwen3.8-max & AA: 53 $\rightarrow$ 40 \newline CritPt: 30 $\rightarrow$ 20 \newline LMArena vision: 1st $\rightarrow$ 3rd \\
\hline
Analyst worker & planning tool chains \newline long agent loops & AA composite \newline AutomationBench-AA & claude-sonnet-5 + gpt-5.6-terra & claude-sonnet-5 + qwen3.7-plus & AA: 42 $\rightarrow$ 26 \newline AutomationBench-AA: 59.6\% $\rightarrow$ 17.4\%$^{{*}}$ \newline (the Anthropic worker is unchanged) \\
\hline
\end{tabular}
\end{table*}

%% file: tables/slots.tex
\begin{table*}
\centering
\caption{The runs of the model-tier test. $k/14$ counts the quantities recovered within
$3\sigma$ of the published values. The gate and review columns give rejections/runs.
Five of the six runs recover at least 10 of the 14 quantities, and
all six reproduce the pattern of
Table~\ref{t:cygx1} (all PDs below the published values, all PAs consistent with them).}
\label{t:slots}
\small
\begin{tabular}{lllrrlrll}
\hline
Run & Assignment & Tools & Hours & US\$ & $k/14$ & mean $|z|$ & Gate & Review \\
\hline
1 & anchor & 0 & 3.5 & 38 & 8/14 & 1.12 & 0/5 & 5/5 \\
2 & anchor & 6 & 21.5 & 121 & 14/14 & 1.24 & 1/20 & 13/19 \\
3 & anchor & 7 & 7.2 & 95 & 14/14 & 1.29 & 1/18 & 11/18 \\
4 & contrast & 8 & 9.5 & 60 & 12/14 & 1.23 & 3/25 & 14/22 \\
5 & contrast & 3 & 6.6 & 102 & 12/14 & 1.23 & 42/49 & 4/7 \\
6 & contrast & 3 & 3.1 & 51 & 12/14 & 1.23 & 2/8 & 3/6 \\
\hline
\end{tabular}
\end{table*}

%% file: appendix.tex
\providecommand{\us}{\_\allowbreak}
\setlength{\emergencystretch}{1.5em}
\renewcommand{\dbltopfraction}{0.9}
\renewcommand{\dblfloatpagefraction}{0.7}

\section{Brief and verification records of the stability tool}
\label{a:toolspec}

 During the stability analysis (Sect.~\ref{ss:res-stability}), the analyst lead wrote the following
brief for the tool \texttt{ixpe\us{}pa\us{}pd\us{}stability}. The tool-writing trials of
Sect.~\ref{ss:trials} use the same brief unchanged. We quote it from the run log up to its third
computation step.

\begin{quote}\footnotesize\raggedright
``Extend \texttt{xspec\us{}utils} with a new pure-numpy/scipy human-half module
\texttt{lib/pa\us{}pd\us{}stability.py} (must NOT import xspec or ixpeobssim), one
\texttt{skills/session.py} wrapper \texttt{ixpe\us{}pa\us{}pd\us{}stability}, and one
\texttt{TOOL\us{}MANIFEST} entry [\ldots]. Signature: \texttt{pa\us{}pd\us{}stability(obs\us{}db\us{}path,
obsids, output\us{}dir, sources=None, variability\us{}p\us{}threshold=0.01, sf\us{}lags\us{}ks=None (default
[5,10,20] ks), min\us{}seg\us{}for\us{}sf=10)}. Per obsid it reads
\texttt{\{obs\us{}db\us{}path\}/\{obsid\}/pcube\us{}timeseg\us{}\{obsid\}.csv} (reuse
\texttt{red\us{}ixpe.\us{}read\us{}existing\us{}timeseg\us{}csv} to obtain provenance + rows\us{}by\us{}segid +
integrated; [\ldots]). Compute: (1) weighted mean \texttt{x\us{}bar=sum(w\us{}i x\us{}i)/sum(w\us{}i)} with
\texttt{w\us{}i=1/sigma\us{}i\^{}2} and \texttt{chi2=sum(w\us{}i(x\us{}i-x\us{}bar)\^{}2)}, \texttt{dof=N-1}, for
PD, QI, UI, and PA; PA uses mod-180 handling (doubled-angle circular weighted mean, residuals
wrapped to \texttt{(-90,+90]} via \texttt{((d+90)\%180)-90}); (2) AUTHORITATIVE 2D Stokes
\texttt{chi2\us{}2D=chi2\us{}QI+chi2\us{}UI}, \texttt{dof\us{}2D=2(N-1)},
\texttt{p\us{}2D=scipy.stats.chi2.sf}; variable flag = \texttt{p\us{}2D<variability\us{}p\us{}threshold};
also per-quantity p-values; (3) moment-estimator intrinsic scatter
\texttt{sigma\us{}int=sqrt(max(0,sigma\us{}obs\^{}2-mean\us{}var))} with \texttt{upper\us{}limit} flag when
the term \texttt{<=0}, for PA (wrapped residuals) and PD; [\ldots]''
\end{quote}

The run log records the following gate, review and commit events for this tool and for its
plotting tool, \texttt{ixpe\us{}pa\us{}pd\us{}stability\us{}plot}. We abbreviate the report and commit hashes.

\begin{quote}\footnotesize\raggedright
\texttt{gate.result: rejected: no\us{}fabrication.vocabulary:}\\
\texttt{\ \ xspec\us{}utils/tests/test\us{}pa\us{}pd\us{}stability.py:12 --- banned phrase `synthetic'}\\[2pt]
\texttt{gate.result: passed (report 1df76135\ldots{} $\rightarrow$ cf4e402e\ldots)}\\
\texttt{review.result: approved. ``The implementation matches the requested PA/PD/Stokes
stability capability [\ldots] the committed tests are adequate for the core science
capability: they include a named reproducibility test on real input, real-data and marked
synthetic capability discrimination, independent chi2 recomputation, CSV-reader differential
coverage [\ldots]''}\\[2pt]
\texttt{commit: ixpe\us{}pa\us{}pd\us{}stability $\rightarrow$ a6e427c5}\\[6pt]
\texttt{review.result: rejected. ``\,`stability\us{}plot' treats a nonexistent `catalog\us{}csv'
as an empty catalog [\ldots] the code silently accepts a missing required catalog CSV and still
reports success [\ldots]''}\\
\texttt{review.result: rejected. ``\,`\us{}load\us{}catalog\us{}rows()' does not validate the CSV
header before returning [\ldots] The catalog is not checked up front for all fields the plotter
later requires [\ldots]''}\\
\texttt{review.result: approved (third submission)}\\
\texttt{commit: ixpe\us{}pa\us{}pd\us{}stability\us{}plot $\rightarrow$ 727fbace}
\end{quote}

 The committed tests write two fixture CSVs with hand-chosen values,
\texttt{synth\us{}pa\us{}pd\us{}variable} and \texttt{synth\us{}pa\us{}pd\us{}constant}, into a
temporary directory. One test checks that the tool flags the first as variable and the second as
constant. The lead wrote this test and told the coder to add it, although it had earlier told the
coder that ``a test never generates data'' (question B.1 in \ref{a:transcripts}). The gate rejected the word
`synthetic' in the test file, and the coder reworded the text. The co-lead then approved the tests,
citing ``marked synthetic capability discrimination''. Neither check stopped the tests from writing these fixture CSVs.

\section{Questions from the coder to the lead}
\label{a:transcripts}

While writing \texttt{ixpe\us{}pa\us{}pd\us{}stability}, the coder asked the lead two questions. We summarize each question and quote the lead's answer from the run log.

\medskip
\noindent\textbf{B.1 Missing synthetic test data.} One check in the brief required two synthetic observations, one with variable and one with constant polarization. The coder found no such data on disk and offered two options. In option 1, the lead or the human would point it to pre-generated files. In option 2, the coder would drop the check and test the discriminating power with real data and arithmetic tests. The lead answered:

\begin{quote}\small
``Go with option 2 --- do NOT fabricate synthetic fixtures. My checks of disk state confirm
your finding: \texttt{data\us{}catalog} lists 16 real local IXPE ObsIDs and no
\texttt{synth\us{}pa\us{}pd\us{}*} entry exists [\ldots] No synthetic fixture was ever pre-generated,
and the policy is firm: a test never generates data, and the toolkit contains no
data-fabricating code. Since the fixtures don't exist, drop the `variable vs constant'
synthetic capability check from this pass and cover discriminating power with real, local data
plus lib-level arithmetic --- exactly as the existing test already does [\ldots]''
\end{quote}

\noindent\textbf{B.2 Conflicting instructions in the brief.} The brief asked for a module that imports neither \texttt{xspec} nor \texttt{ixpeobssim}. It also asked the module to reuse a CSV reader from \texttt{red\us{}ixpe}, which imports both at module level. The coder noticed this conflict and proposed its own CSV reader. The lead agreed, because importing \texttt{red\us{}ixpe} ``would pull the entire science
stack into your pure-numpy/scipy lib --- exactly what the brief forbids.''

\section{Result records of the analyses}
\label{a:results}

\begin{figure}
\centering
\includegraphics[width=0.99\linewidth]{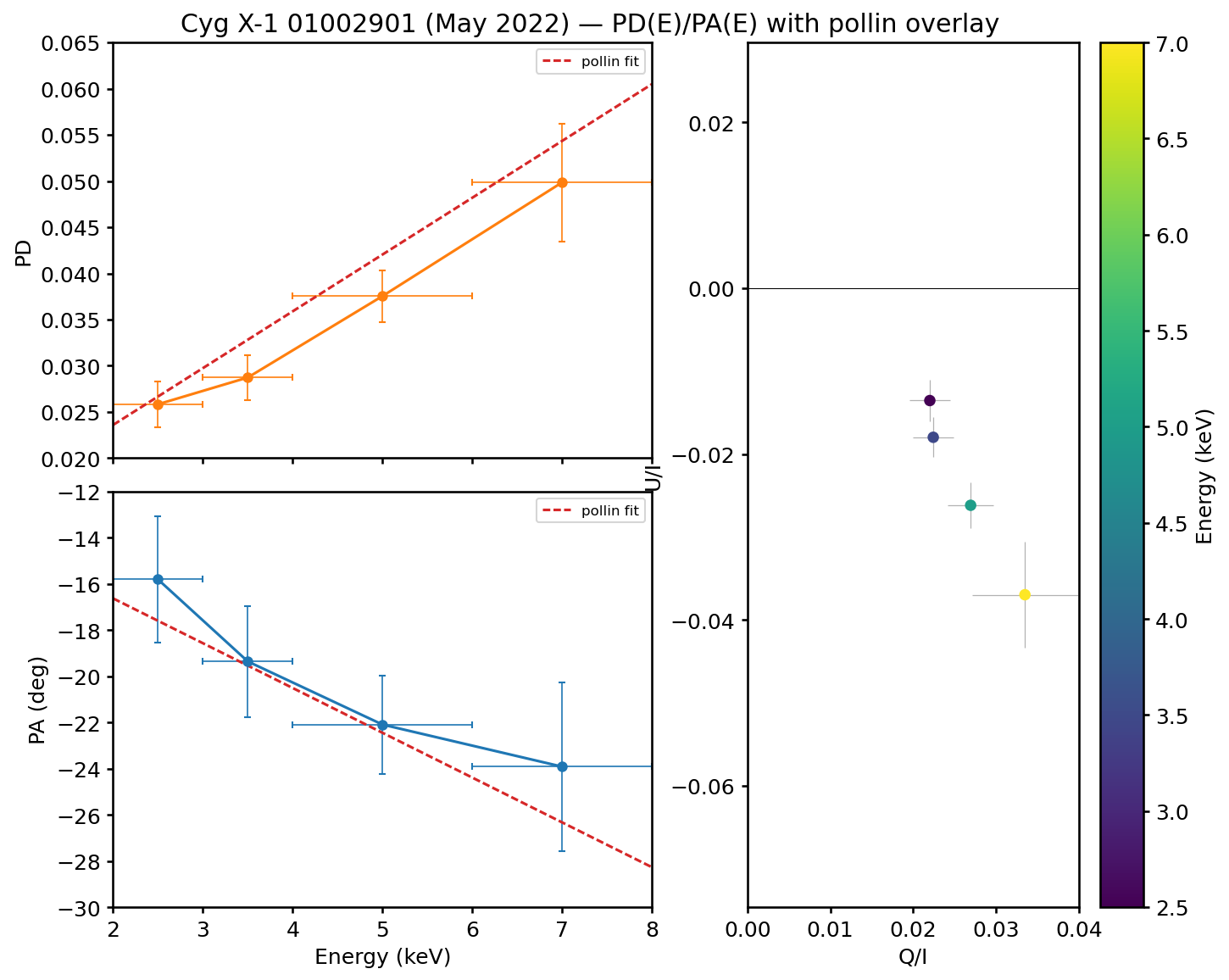}
\caption{PD (as a fraction, upper left) and PA (lower left) against energy for Cyg X-1 observation 01002901 (May 2022) of the reproduction analysis (Sect.~\ref{ss:res-calibration}). Dashed line: the fitted XSPEC \texttt{pollin} model. Right: $u$ against $q$, coloured by energy. Produced by the tool \texttt{ixpe\us{}pcube\us{}plot}.}
\label{f:cygx1figs}
\end{figure}

\begin{figure*}
\centering
\includegraphics[width=\textwidth]{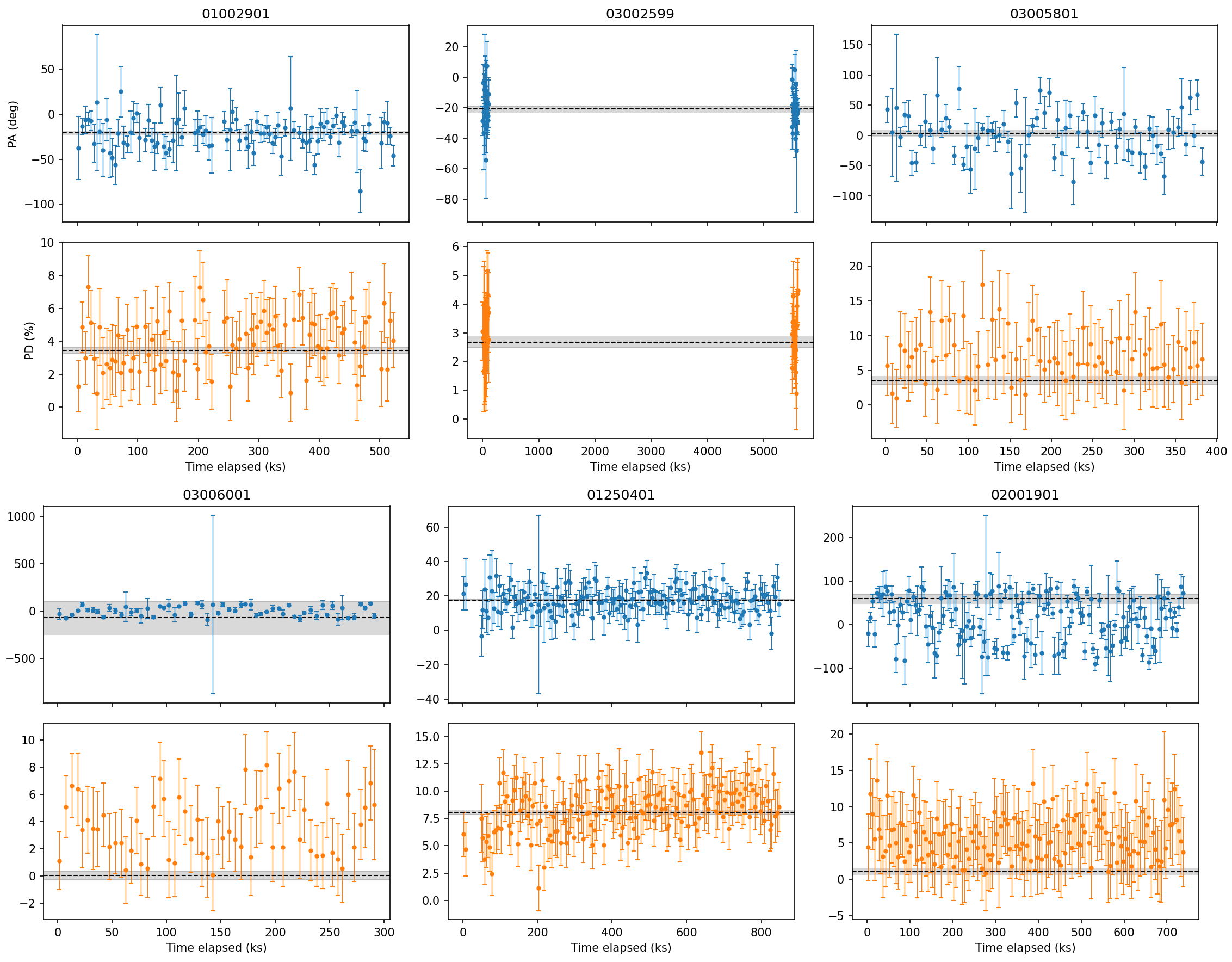}
\caption{PA (upper panels) and PD (lower panels) of the 5~ks segments of the six observations of the stability analysis (Sect.~\ref{ss:res-stability}). Top row: Cyg X-1 01002901, Cyg X-1 03002599 and Swift J1727.8$-$1613 03005801; bottom row: Swift J1727.8$-$1613 03006001, 4U 1630$-$47 01250401 and LMC X-1 02001901. Observation 03002599 consists of two visits 62 days apart. Dashed line and grey band: PA or PD of the whole observation and its $1\sigma$ uncertainty. Produced by the tool \texttt{ixpe\us{}pa\us{}pd\us{}stability\us{}plot} (\ref{a:toolspec}), which draws the six observations in one row; we arranged them in two rows.}
\label{f:stability}
\end{figure*}

\begin{figure*}
\centering
\includegraphics[width=0.95\textwidth]{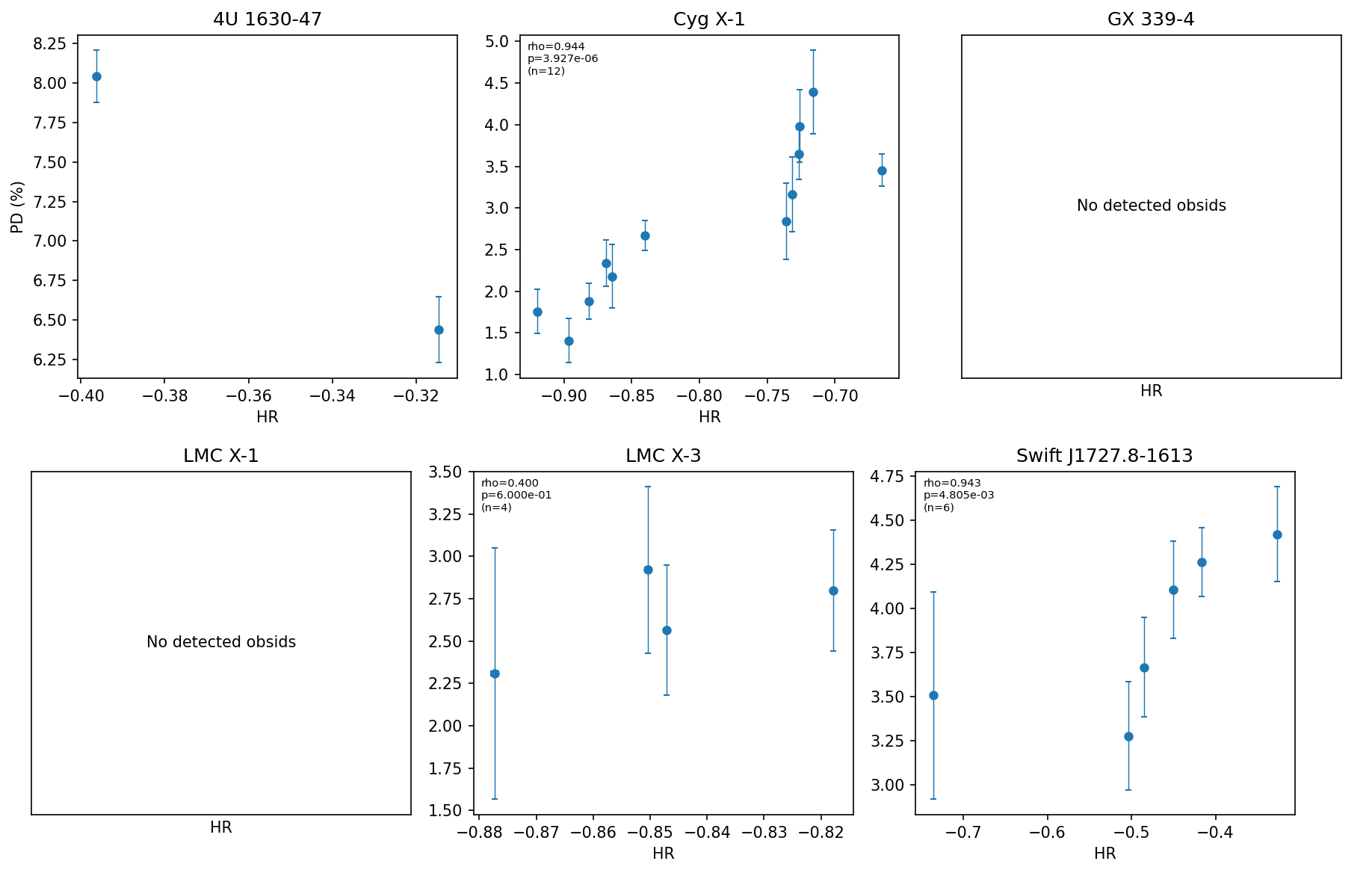}
\caption{PD against HR for each source of the survey, observations with a detection only, produced by the tool \texttt{pd\us{}hr\us{}survey\us{}plot}. Panels of sources with at least four detections give the Spearman coefficient, its $p$-value and the number of detections. GX 339$-$4 and LMC X-1 have no detections. The tool draws the six panels in one row; we arranged them in two rows.}
\label{f:pdhrsource}
\end{figure*}

\begin{figure}
\centering
\includegraphics[width=0.8\linewidth]{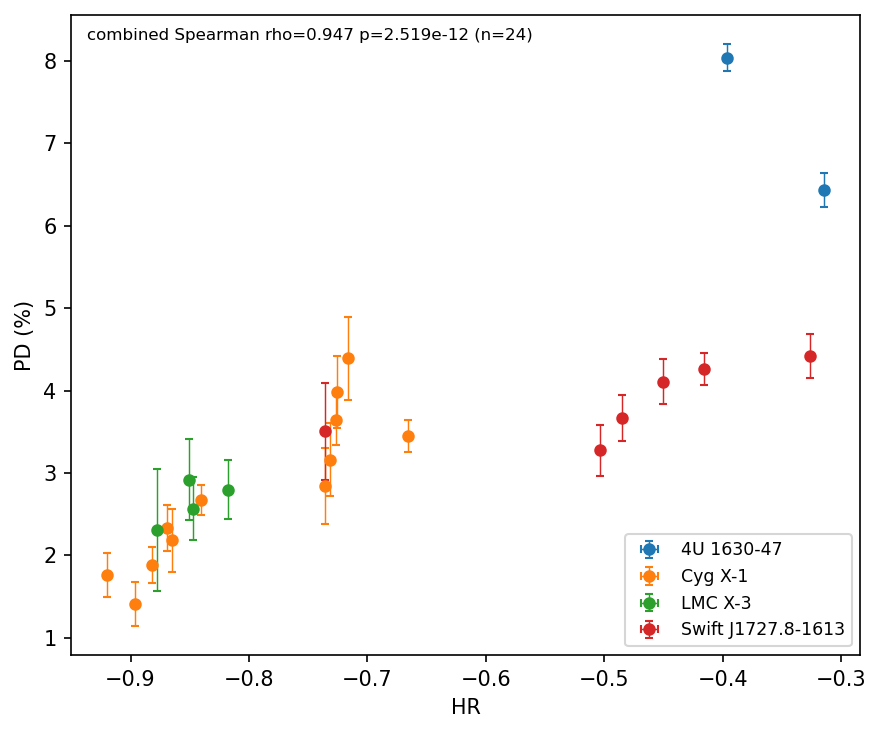}
\caption{PD (2--8 keV) against HR for the 24 observations of the survey with a detection (PD above MDP99), coloured by source, produced by the same tool as Fig.~\ref{f:pdhrsource}. The five observations without a detection (HR from $-0.97$ to $-0.87$) are not shown.}
\label{f:pdhr}
\end{figure}

\section{The $q$--$u$ time-lag analysis}
\label{a:qulag}

This appendix gives the details of the fourth analysis (Sect.~\ref{ss:res-qulag}). The analysis has
no published reference and no expected detection. A false detection is therefore its main risk. The request stated that ``a null result (no significant lag), or per-observation upper
limits, is the expected, correct, and publishable outcome'', and it instructed the agents not to
``tune the null, hand-roll the correlation, or back-fill a detection''.

\emph{Tool writing.} Three tools were already in the toolkit: the slicer
\texttt{ixpe\us{}qu\us{}slice\us{}pcubes}, the DCF tool \texttt{ixpe\us{}qu\us{}lag\us{}dcf}, and the
plotting tool \texttt{ixpe\us{}qu\us{}lag\us{}dcf\us{}plot}. One analyst worker ran all three before
the first coding session. Seven coding sessions then changed these tools or added a new one, and
five of them committed. Two commits fixed the handling of numerical arguments in the slicer and the DCF tool. One stopped the slicer from splitting a merged 386~ks observation (04002899) at each of its orbital gaps. The other two added the absolute lag of the DCF peak, $|\tau_{\rm peak}|$, to the Spearman tests across the segments and to the plotting tool. Each commit passed the gate
and the review on its first submission. The sixth session worked on a further change to the slicer. The gate
rejected it once for banned vocabulary, and the session ended before review. The seventh session
wrote a new tool, \texttt{ixpe\us{}pol\us{}report}, for a table of the polarization of each
observation. The gate rejected it once for banned vocabulary, and the review rejected it four
times. In total, the gate ran 11 times and rejected two submissions, and the review ran nine times
and rejected four.

\emph{Shortfalls.} The final report of the run states two shortfalls. First, the report has no measurement for the two proprietary observations of Sect.~\ref{ss:res-qulag}, SS~433 (04004001, 745~ks) and GS~1354$-$64 (05250399, 452~ks). They are two of the long observations that the
request expected to anchor the analysis. Second, the report lacks the separate polarization table
that the request asked for. The last review of \texttt{ixpe\us{}pol\us{}report} found that ``the
resumable checkpoint path can silently return an invalid successful science row after an
interrupted write'', and the lead started no further session for this tool. The report therefore
uses the polarization significance computed inside the DCF tool, and it says so.

\emph{Result.} The permutation probabilities of the 11 segments without a significant correlation range from 0.065 to 0.933. The per-test criterion of the request requires $p<0.01$ and a 68\% interval of the peak lag $\tau_{\rm peak}$ that excludes zero. The candidate in Cyg~X-3 observation 02250301 meets this criterion. Its interval, from flux randomization alone \citep{peterson1998}, is $[-100.94, -57.68]$~ks. However, \citet{peterson1998} combine flux randomization with random-subset
selection, which tests the sensitivity to the sampling, and find that either method alone generally
gives smaller uncertainties. The interval of the candidate is therefore probably too narrow.

The pipeline also ran four Spearman tests across the segments, and two of them have $p<0.01$. The signed
$\tau_{\rm peak}$ correlates with the polarization significance ($\rho = +0.776$, $p = 0.0030$),
and $|\tau_{\rm peak}|$ anticorrelates with a brightness proxy ($\rho = -0.734$, $p = 0.0065$). The
report claims neither result. It attributes the first correlation to the three segments with the
weakest polarization detections, which all have large negative $\tau_{\rm peak}$. Their peaks lie
at lags of 0.4 to 1.6~Ms, longer than a single visit. The DCF tool searches lags up to half the
time span of a segment, so for a segment with several visits it also searches lags between visits.
The report recommends restricting the peak search in future runs. It attributes the second
correlation to the sampling. Fainter sources need wider slices and were observed over several
visits, so both the shortest and the longest lag that can be measured grow as brightness falls.

\emph{What the checks did.} The reviewer read the DCF code, compared it with the noise-corrected normalization in Eq.~(3) of
\citet{edelson1988}, and accepted the report. The reviewer also
wrote that this reference ``explains why few-pair lag bins yield $|\mathrm{DCF}|>1$''. The
reference does not support this. It gives the expected scatter of the DCF of uncorrelated series,
about $\pm M^{-1/2}$ in a lag bin of $M$ pairs, and does not discuss $|\mathrm{DCF}|>1$. For the
peaks of $|\mathrm{DCF}| = 2.2$--$3.4$, the report and the reviewer based the significance on the
permutation probability alone. One analyst worker computed all results, and the second worker only
plotted them again. The analyst lead contacted the human 11 times during the run. Ten contacts
discussed the design of the tools to be written, and one asked whether to accept the 45 readable
observations as the sample.

\emph{Cost.} The run made 613 model calls and 1{,}592 tool calls. The coding agent accounts for
78\% of the cost in Table~\ref{t:campaign}, in seven tool-writing sessions.

\section{Implementation and cost}
\label{a:impl}

\begin{table}
\centering
\caption{Models in each role, from the run records of each analysis; the coder's model comes from the model configuration at the time of the run. A dash marks a role that the pipeline did not yet have. Table~\ref{t:seats} gives the models of the runs of Sect.~\ref{s:generality}, and Table~\ref{t:sweepevents} those of the tool-writing trials of Sect.~\ref{ss:trials}.}
\label{t:roster}
\scriptsize
\setlength{\tabcolsep}{2pt}
\begin{tabular}{@{}L{0.16\columnwidth}L{0.26\columnwidth}L{0.26\columnwidth}L{0.26\columnwidth}@{}}
\hline
Role & Sect.~\ref{ss:res-calibration} & Sects.~\ref{ss:res-stability}--\ref{ss:res-survey} &
  Sect.~\ref{ss:res-qulag} \\
\hline
Reviewer & \texttt{claude-fable-5} & \texttt{claude-opus-4-8} & \texttt{claude-fable-5} \\
Analyst lead & \texttt{claude-opus-4-8} & \texttt{claude-opus-4-8} & \texttt{claude-opus-5} \\
Co-lead & \texttt{gpt-5.5} & \texttt{gpt-5.5} & \texttt{gpt-5.6-terra} \\
Analyst workers & \texttt{claude-sonnet-4-6}\newline\texttt{gpt-5.5} &
  \texttt{claude-sonnet-4-6}\newline\texttt{gpt-5.5} &
  \texttt{claude-sonnet-5}\newline\texttt{gpt-5.6-terra} \\
Coder & \texttt{claude-sonnet-5} & \texttt{claude-sonnet-5} & \texttt{claude-sonnet-5} \\
Librarian & \texttt{claude-sonnet-4-6} & \texttt{claude-sonnet-4-6} & \texttt{claude-sonnet-5} \\
Status monitor & --- & --- & \texttt{gpt-5.6-luna} \\
\hline
\end{tabular}
\end{table}

Table~\ref{t:roster} lists the models of each analysis.

{\sloppy \pipeline\ has 34{,}289 lines of Python in 121 modules. The typed messages between agents (Sect.~\ref{ss:team}) are 35 Pydantic models. The deterministic test suite has 1{,}824 tests in 27{,}444 lines. The benchmark harnesses of Sect.~\ref{s:verification} have 16{,}716 lines and 499 tests. \par}

On the working branch, \toolkit\ holds three tool packages with 25 tools. The \texttt{lib/} halves of these tools (Sect.~\ref{ss:design}) have 8{,}247 lines, the \texttt{skills/} halves 3{,}444 lines, and the tests of the toolkit 11{,}130 lines. No file in the toolkit imports \pipeline. Thus a human can reproduce a tool with only the science software and the toolkit.

\begin{sloppypar}
The 35 phrases of the banned-vocabulary scan (Sect.~\ref{ss:selfext}) are example data, sample data, fake data, fake value, fake result, mock data, dummy data, made up, made-up, fabricated, fabricate, fictional, fictitious, hardcoded result, hard-coded result, canned, pretend, bogus, toy data, representative data, representative value, representative result, for now, in operational mode this would, in production this would, in practice this would, would actually, would normally, for demonstration, for illustration, placeholder, TODO, FIXME, XXX, HACK.
\end{sloppypar}

\pipeline\ records the tokens and the cost of each model call. The experiments reported here cost about US\$2{,}200 at list prices. This total comprises the analyses of Table~\ref{t:campaign}, US\$259 for the benchmark of Sect.~\ref{ss:catchrate}, US\$984 for the tool-writing trials of Sect.~\ref{ss:trials} and US\$466 for the runs of Sect.~\ref{s:generality}.